\documentclass[12pt,preprint]{aastex}

\usepackage{graphicx}
\usepackage{txfonts}
\usepackage{natbib}
\usepackage{subfigure}
\graphicspath{{./figs/}}
\usepackage{multirow}   
\usepackage[colorlinks,linkcolor=blue,anchorcolor=blue,citecolor=blue]{hyperref}
\usepackage[hyperref]{backref}
\usepackage{rotating}
\usepackage{lscape}
\usepackage[usenames,dvipsnames]{xcolor}

\def\stacksymbols #1#2#3#4{\def\theguybelow{#2}
        \def\verticalposition{\lower#3pt}
        \def\spacingwithinsymbol{\baselineskip0pt\lineskip#4pt}
        \mathrel{\mathpalette\intermediary#1}}
\def\intermediary #1#2{\verticalposition\vbox{\spacingwithinsymbol
        \everycr={}\tabskip0pt
        \halign{$\mathsurround0pt#1\hfil##\hfil$\crcr#2\crcr
                \theguybelow\crcr}}}

\def\gta{\stacksymbols{>}{\sim}{2.5}{.2}}

\def\Msun{M_\odot}

\tabletypesize{\footnotesize}

\begin{document}


\title{X-Ray Properties of AGN in Brightest Cluster Galaxies. I. A Systematic Study of the {\sl Chandra} Archive in the 
$0.2<z<0.3$ and $0.55<z<0.75$ Redshift Range}

\author{Lilan Yang\altaffilmark{1,\dag}, Paolo Tozzi\altaffilmark{2,1}, Heng Yu\altaffilmark{1,\ddag}, Elisabeta Lusso\altaffilmark{3},
Massimo Gaspari\altaffilmark{4,*}, Roberto Gilli\altaffilmark{5}, Emanuele Nardini\altaffilmark{2} \& Guido Risaliti\altaffilmark{6,2} }


\altaffiltext{1}{Department of Astronomy, Beijing Normal University, Beijing, 100875, People's Republic of China}
\altaffiltext{2}{INAF - Osservatorio Astrofisico di Arcetri, Largo E. Fermi, I-50122 Firenze, Italy}
\altaffiltext{3}{Centre for Extragalactic Astronomy, Department of Physics, Durham University, South Road, Durham, DH1 3LE, UK}
\altaffiltext{4}{Department of Astrophysical Sciences, 4 Ivy Lane, Princeton University, Princeton, NJ 08544, USA}
\altaffiltext{5}{INAF - Osservatorio di Astrofisica e Scienza dello Spazio di Bologna, via Gobetti 93/3 - 40129 Bologna - Italy}
\altaffiltext{6}{Universit\'a di Firenze, Dip. di Fisica e Astronomia, via G. Sansone, I-50019 Sesto Fiorentino, Firenze, Italy}
\altaffiltext{\dag}{{\it email: yang\_lilan@mail.bnu.edu.cn} }
\altaffiltext{\ddag}{{\it email: yuheng@bnu.edu.cn} }
\altaffiltext{*}{{\it Einstein} and {\it Spitzer} Fellow}

\begin{abstract}
\noindent
We present a search  for nuclear X-ray emission in the brightest cluster galaxies (BCGs) of a sample of groups 
and clusters of galaxies extracted from the {\sl Chandra} archive. 
The exquisite angular resolution of {\sl Chandra} allows 
us to obtain robust photometry at the position of the BCG, and to firmly identify unresolved X-ray emission when present, 
thanks to an accurate characterization of the extended emission at the BCG position.
We consider two redshift bins ($0.2<z<0.3$ and $0.55<z<0.75$) and analyze all the clusters observed by {\sl Chandra} 
with exposure time larger than 20 ks. 
Our samples have 81 BCGs in 73 clusters and 51 BCGs in 49 clusters in the low- and high-redshift bin, 
respectively.  X-ray emission in the soft (0.5-2 keV) or hard (2-7 keV) band is detected only in 14 and 9 BCGs 
($\sim 18$\% of the total samples), respectively. The X-ray photometry shows 
that at least half of the BCGs have a high hardness ratio, compatible with significant intrinsic absorption.  
This is confirmed by the spectral analysis with a power-law model plus intrinsic absorption.
We compute the fraction of X-ray bright BCGs above a given hard X-ray luminosity, 
considering only sources with positive photometry in the hard band (12/5 sources in the low/high-{\sl z} sample). 
In the $0.2<z<0.3$ interval the hard X-ray luminosity ranges from $10^{42}$ to $7 \times 10^{43}$ erg  
s$^{-1}$, with most sources found below $10^{43}$ erg s$^{-1}$.  In the $0.55<z<0.75$ range, we find a 
similar distribution of luminosities below $\sim 10^{44}$ erg s$^{-1}$, plus two very bright sources of a few 
$10^{45}$ erg s$^{-1}$ associated with two radio galaxies.  We also find that X-ray luminous BCGs 
tend to be hosted by cool-core clusters, despite the majority of cool cores do not host nuclear X-ray emission.
This work shows that our analysis, when extended to the entire {\sl Chandra} archive, can provide a sizable
number of sources allowing us to probe the evolution of X-ray AGN in BCGs as a function of the cosmic epochs.    
\end{abstract}



\section{Introduction} \label{s:intro}
\noindent
Brightest cluster galaxies (BCGs) are defined as galaxies that spend most of their life at the bottom of the potential 
wells of massive dark matter halos.  This particular location favors accretion from satellite galaxies or from gas cooling 
out of the hot phase of the intracluster medium (ICM).  In turn, cooling gas may feed several star formation episodes 
\citep[e.g.,][]{2017Bonaventura} or mass growth of the central super massive black hole (SMBH). 
Therefore, their evolution is directly linked to the dynamical history of the host cluster and to the cycle of baryons in 
cluster cores. For these reasons, BCGs are the largest and most luminous ones among the cluster galaxy population.  
Due to the hierarchical process of structure formation, in dynamically young clusters or major mergers there may be more 
than one BCG, so that BCGs may not be unambiguously defined as the brightest galaxies.  In addition, in such dynamically 
disturbed halos, their position may not coincide with the center of the X-ray emission \citep[see][]{2016Rossetti}.  Despite 
the intrinsic difficulty in defining a unique BCG at any epoch during the lifetime of a virialized massive halo, 
in most of the cases, BCGs are by far the most luminous galaxies in the optical band, and their position is almost coincident 
with the peak of the X-ray brightness, with a typical displacement of less than 10 kpc \citep{2003Katayama}.  This is a 
typical case in relaxed, cool-core cluster, where their identification is straightforward. Alternatively, off-centered or multiple BCGs
\citep[reported for a fraction ranging from 5\% to 15\%, see][]{1999Crawford,2015Hogan} are often associated with signatures of 
ongoing or recent major mergers.

An important property of BCGs is the ubiquitous presence of significant nuclear radio emission.  \citet{Best:2005,Best:2007} 
showed that BCGs are more likely to host a radio-loud AGN by a factor of several with respect to normal ellipticals, although only
20-30\% of the BCGs can be defined radio-loud AGN. The likely cause of this behavior is the increasing amount of fueling 
surrounding the BCG in the form of cold gas cooling out the hot phase. Indeed, the dependence on the SMBH mass of the 
radio-loud AGN fraction \citep{Best:2005} mirrors that of the cooling rate from the hot halos.
Recently, $^{13}$CO line emission from molecular gas has been detected in BCGs \citep[see, e.g.,][]{2017Vantyghem}.

The origin of the cold gas is problematic, to say the least.  Observationally, pure isobaric cooling flows \citep{1994Fabian} are 
not observed, and X-ray spectra indicate that they must be suppressed in flux at least  by a factor of 10-100, in particular in the 
soft X-rays (e.g., \citealt{2006Peterson}). Since star formation is linked to the cooling gas, BCG star formation rates in the range 
1-100 $M_\odot yr^{-1}$ are observed to be quenched as well, albeit with large scatter that is due to the temporal delays involved 
(e.g., \citealt{2016Molendi}).  The thermal structure of the hot gas and the quenched star formation can be reconciled 
by invoking a plethora of complex phenomena collectively named feedback, where AGN are the most likely feedback agent.  
The AGN present in the central galaxies can inject mechanical energy through relativistic jets or winds.  This energy is most likely 
thermalized at $r \gta10$\,-\,100 kpc via buoyant hot bubbles, weak shocks, and turbulence 
(e.g., \citealt{2013Gaspari,Barai:2016}).  While the macro imprints of AGN feedback can be resolved by current X-ray 
telescopes, the actual micro carrier of kinetic energy is still debated.  The radio electron synchrotron power of relativistic jets 
is typically over 100 times lower than the total cavity power (\citealt{McNamara:2012}); the {\it Fermi} telescope has not detected 
any substantial gamma-ray emission within bubbles, hence excluding relativistic protons; in addition, several (`ghost') cavities 
have been found to be devoid of radio emission \citep[see][]{2004Birzan}. Another source of feedback can be massive 
sub-relativistic outflows, typically with a wider opening angle compared to jets, which are able to entrain the background gas 
along the path. Detections of multiphase AGN outflows are booming during the past few years 
(e.g., \citealt{Tombesi:2013,Russell:2014,Combes:2015,Feruglio:2015,Morganti:2015_rev}). 
Overall, the radio power can be a tracer of feedback, although there are also other mechanical injection channels that are 
not necessarily associated with an increase of nuclear radio power. It is thus best to refer to this mode of feedback 
as the {\it 
mechanical} mode (which includes both relativistic jets and sub-relativistic outflows) instead of as the radio mode.

In recent years, a detailed picture of AGN feeding in massive halos has emerged
(e.g., \citealt{Gaspari:2013_CCA, Voit:2015_ge, Voit:2015_nat}). According to this picture, warm filaments 
and cold clouds are expected to condense out of the hot gaseous halo of the massive galaxy, group, or cluster in a 
multiphase condensation cascade and rain toward the central AGN. Inelastic collisions promote 
angular momentum cancellation, boosting the accretion rate and thus increasing the nuclear AGN power.  This mechanism 
is known as chaotic cold accretion (CCA).  The CCA feeding triggers the feedback via 
AGN jets or outflows in a tight self-regulated loop \citep[see][]{Gaspari:2017_CCA}.

This is a promising mechanism, since, on the basis of the ubiquitous observations of a quenched cooling rate in cool cores, 
the mechanical mode of AGN feedback is expected to be tightly self-regulated in 
most -- if not all -- BCGs (e.g., \citealt{2009Sun}).  This mode is often associated with radiatively inefficient accretion on AGN 
\citep{2012Fabian}.  However, a fraction of BCGs also shows substantial X-ray emission, suggesting the coexistence of a 
radiatively efficient accretion disk or momentarily boosted rain near the inner SMBH hosted by the BCG.  The X-ray properties of 
BCGs have been systematically investigated by \citet{2013Russell} in a low-redshift sample, to explore the relation between 
nuclear X-ray emission and AGN cavity power.  They found that half of their sample has detectable unresolved X-ray 
emission.  They estimated the accretion rate from the cavity power (assuming some efficiency), finding  that the nuclear radiation 
exceeds the mechanical power when the mean accretion rate is above a few percent 
of the Eddington rate ($\sim 22 \, \Msun$\,yr$^{-1}$ for a $10^9\,\Msun$ SMBH), marking the transition from radiatively 
inefficient AGN to quasars, as expected from the fundamental plane of black hole activity \citep{2003Merloni}.  
As before, they remarked that cold gas fueling is the likely source of accretion 
\citep[e.g.,][]{1986Nulsen,2005Pizzolato,2016McNamara}.
\citet{2013aHlavacek-Larrondo} investigated the nuclear X-ray emission of BCGs in bright X-ray clusters with clear X-ray 
cavities.  They found a  strong evolution in their nuclear X-ray luminosity, at least by a factor of $\sim$\,10 in the 
$0<z<0.6$ redshift range, speculating that the 
transition from  mechanically dominated AGN to quasars occurs at high redshift for the majority of the massive cluster 
population.

The analysis of both \citet{2013Russell} and \citet{2013aHlavacek-Larrondo} are based on a sample of BCGs whose host 
cluster shows large X-ray cavities in the ICM.  The presence of cavities, together with a nuclear radio 
power, allows one to estimate the mean accretion rate onto each galaxy on a time scale of $\sim$\,$10^8$ years.  Here we 
relax this requirement to extend the investigation of unresolved X-ray emission from BCGs to any virialized halo, defined
by the presence of diffuse emission from its ICM. Clearly, with these selection criteria, we are dominated by halos with low 
X-ray surface brightness, and therefore we are not able to search for X-ray cavities.  Our long-term plan is to
collect enough archival, multiwavelength data to use SMBH mass estimate and properties of the environment (such as mass of the 
host halo, cool-core strength, presence of cavities, and dynamical state of the halo) with the final goal of exploring the origin 
of the X-ray emission, 
the accretion regime in BCGs at different epochs and environments, and the origin of the feeding gas and obscuring material 
around the SMBH.  In this first paper of a series, our immediate science goal is to assess our capability of tracing the X-ray 
properties of the BCGs across the wide range of groups and clusters of galaxies currently available in the  {\sl Chandra} archive.  
In particular, we focus on the 2-10 keV nuclear luminosity of BCGs at two different cosmic epochs.  Only the exquisite angular 
resolution of {\sl Chandra} data allows us to unambiguously identify the presence of unresolved X-ray emission embedded in the 
much brighter thermal ICM emission, which must be efficiently modeled and subtracted. 

The paper is organized as follows. In \S 2 we describe the sample selection.  In \S 3 we describe the data reduction and 
analysis, and in \S 4 we provide the results of the X-ray properties of BCGs and the correlation between X-ray and radio 
nuclear emission, and the link with the cool-core strength.  In \S 5 we discuss the possible implications for AGN feeding 
and feedback that can be obtained from our study.  Finally, in \S 6 we summarize our conclusions.  
Throughout the paper, we adopt the seven-year WMAP cosmology ($\Omega_{\Lambda} =0.73 $, $\Omega_m =0.27$, 
and $H_0 = 70.4 $ km s$^{-1}$ Mpc$^{-1}$ \citep{2011Komatsu}.  Quoted error bars  correspond to a 1 $\sigma$ 
confidence level unless noted otherwise.

\section{Sample definition}

\subsection{X-Ray Data Selection}
\noindent
To achieve our science goals, we aim at considering both cool-core and non-cool-core clusters, with no preselection based on 
cluster properties, except for the firm detection of extended ICM thermal emission, which is the unambiguous signature
of a virialized halo.
Therefore we initially consider the entire {\sl Chandra} ACIS archival observations listed under the category "clusters of 
galaxies".  This maximally inclusive selection simply aims at collecting the largest number of BCGs imaged with the best
angular resolution.  In fact, the vast majority of {\sl Chandra} ACIS aimpoints coincide with the cluster center, ensuring the best
angular resolution at the BCG position and therefore allowing us to identify unresolved emission above the
level of the surrounding ICM.  This aspect is key to our research strategy, since the capability of 
detecting unresolved emission embedded in the ICM is rapidly disappearing as the point spread function is degraded as
a function of the off-axis angle.  
We are aware that the large source list initially selected in this way does not constitute a complete sample.  In addition, this 
choice does not allow any control on possible selection bias.  {\bf On the other hand, due to the intrinsic differences among
cluster samples with difference selection \citep[see][for X-ray and Sunyaec-Zel'dovich, SZ, selected clusters samples]{2017Rossetti}, 
a complete and unbiased sample of virialized halos will necessarily be a mix of clusters selected with different criteria.  This 
consideration pushes us to exploit the entire {\sl Chandra} archive with no further restrictions, as an acceptable proxy to 
an unbiased cluster sample.}  Our plan is to test our strategy and eventually extract well-defined 
subsamples from the main parent sample after completing the collection of useful X-ray data. 

Since we wish to explore the X-ray properties of BCGs as a function of the cosmic epoch, we first apply our method in  two 
redshift bins that include a sufficiently large number of clusters (i.e.,  at least 50 in each of them). 
When the same target has multiple exposures, we decide to choose the ObsIDs with the largest total exposure between ACIS-I 
and ACIS-S, and avoid combining the two detectors for simplicity.  In addition, we discard short observations if taken in an 
observing mode different from the bulk of the observations. In this work, aiming essentially in testing our strategy, 
we choose to analyze all the groups and clusters observed with total exposure time $t_{exp}> 20 $ ks to ensure a good
characterization of the extended ICM emission, and eventually perform a spatially 
resolved analysis of the surrounding ICM whenever possible.

In defining the low-redshift bin, we prefer to avoid nearby clusters, so that we can always sample the background from the ICM-free regions around the clusters in the $8\times 8$ arcmin field of view (corresponding to one {\sl Chandra} ACIS CCD).  
We find that the 
choice $0.2<z<0.3$ allows us to obtain a sufficiently large sample and also have a few sources overlapping with the sample of
\citet{2013Russell} for a direct comparison.  We choose the range $0.55<z<0.75$ for the high-redshift bin to include a 
sizable number of clusters.  Moreover, with these choices, we probe a redshift range comparable to that
explored by \citet{2013aHlavacek-Larrondo}.  We have 73 and 49 clusters in the low- and high- redshift bin, respectively.\footnote{SC 1324+3051 
and SL J1634.1+5639 
are removed from the low-redshift bin, since they do not show any 
X-ray extended emission and therefore their virialization status is uncertain.   We also removed CODEX53585,
SC 1604+4323 and RCS 1325+2858 from the high-z bin for lack of visible extended emission.} 
With this choice we aim at delivering a first investigation of the typical X-ray luminosity of BCGs in virialized halos
on a time scale of about 3 Gyr (from $\langle z\rangle = 0.65$ to $\langle z \rangle = 
0.25$), paving the way to an eventual  comprehensive study based on the entire  {\sl Chandra} archive. 

\subsection{BCG Identification}
\noindent
As we discussed in the Introduction, a BCG can be defined as a galaxy that spent a significant part of its life at the
center of a large dark matter virialized halo.  This opens up the possibility that each group or cluster hosts more than 
one BCG at a given time.  Or, more likely, that at any time, it is possible to identify one or more past-BCGs, and at least one 
current BCG. A complete BCG identification strategy based on these premises is beyond our reach with present 
data, and we are necessarily restricted to those galaxies that are currently experiencing their BCG phase. 
Therefore, we proceed first by identifying the BCG in the optical band among those with a redshift 
(when available) that is compatible with the cluster redshift, starting our search from the maximum of the cluster X-ray emission. 
In most cases, we rely on previous identification of the BCG published in the literature. Then, we search for galaxies 
that have been identified as secondary BCGs in the literature, if any.  We do not apply further criteria for the identification 
of the BCG.  Therefore, we also need to collect high-quality multiwavelength data for the same fields selected in the X-ray band.  We make 
use of {\sl Hubble Space Telescope (HST)} images or other lower quality optical data whenever available.
In the worst cases, when there are no records in the literature, or no HST images, we have to rely on the best 
information we can recover from the NASA Extragalactic Database\footnote{https://ned.ipac.caltech.edu/.}.  In this case, 
we avoid searching for a secondary BCG. 

In detail, we obtain the most accurate position of the nucleus of the BCG in the following 
way.  First, we inspect the X-ray image, and obtain the coordinates at the maximum of the X-ray surface brightness 
emission, identified with {\tt ds9} in the total band (0.5-7 keV) image.  In the case of very smooth X-ray emission, 
we choose the emission-weighted center.  We stress that the initial choice of the X-ray center does not affect the final BCG 
identification, since it is used merely as a starting point.  Eventually, we search for HST images within 2 arcmin from the 
approximate X-ray center.  We download an optical image from the HST 
archive, \footnote{https://archive.stsci.edu/hst/search.php.} visually inspect the X-ray and optical images, and finally select 
the position of the nucleus of the BCG.  In addition, we search for the BCG position in the literature from different works 
to confirm our BCG identification.  When no HST data are 
available, we refer to the literature and/or to the NASA Extragalactic Database, where we search for the 2MASX or SDSS 
counterpart closest to the X-ray center we preselected. The positional accuracy obtained in this way is always on the order 
of $\sim$\,1  arcsec, which is sufficient to unambiguously identify the X-ray unresolved emission associated with the BCG, 
when present.  In the cases of clear unresolved X-ray emission associated to the BCG, we slightly refine the center of 
the extraction region to sample at best the BCG X-ray emission.  In all the other cases (no unresolved emission), the typical 
$\sim 1$ arcsec uncertainty on the position of the BCG nucleus has a negligible impact on the estimation of the upper limit
to the BCG X-ray emission.

\begin{figure}
\begin{center}
\includegraphics[width=12cm]{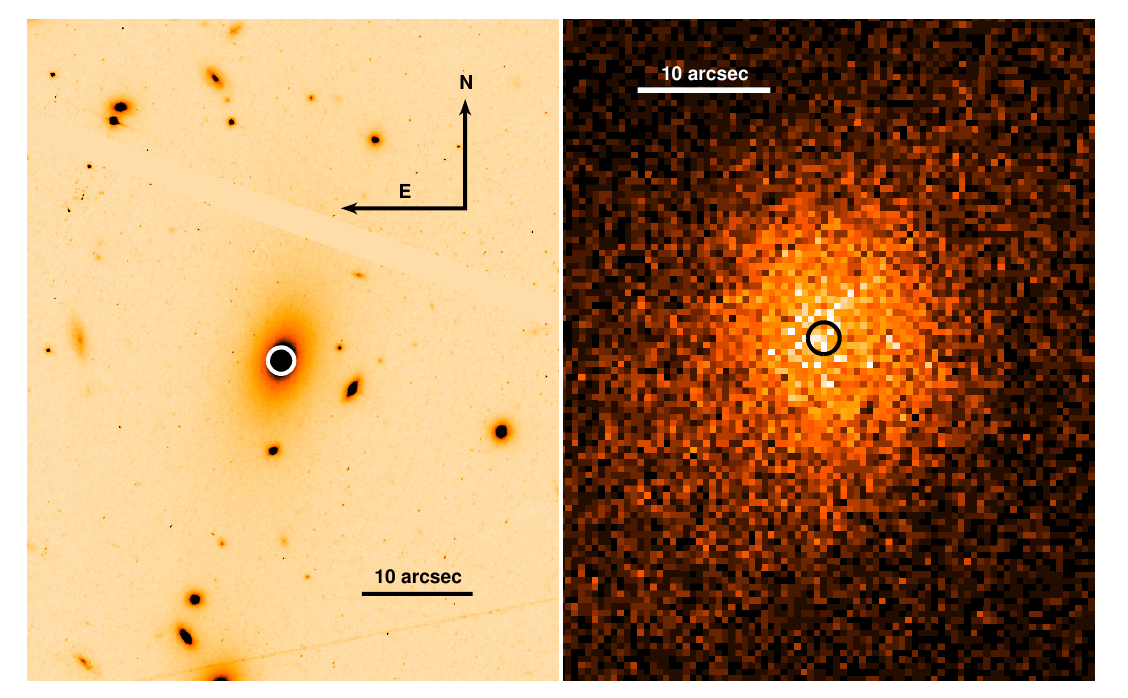} 
\includegraphics[width=12cm]{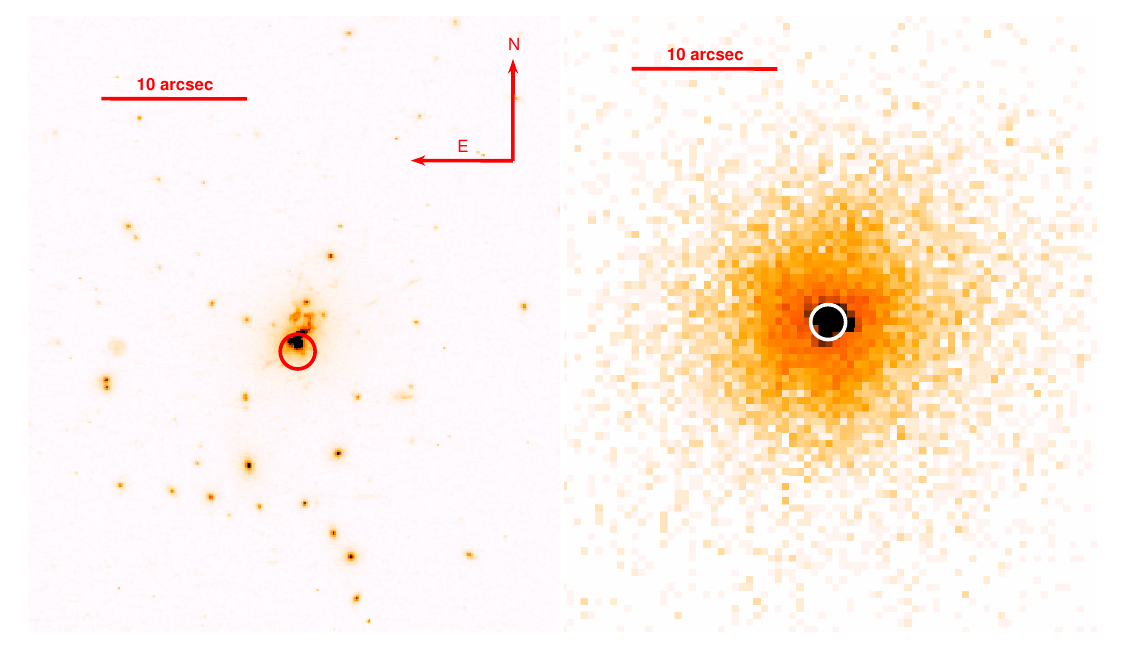}  
\caption{\label{HST_images}Upper panels: HST/ACS image (left) and {\sl Chandra} hard band image (right) of 
MS 0735.6+7421 at $z=0.216$.  The BCG position, taken from the optical image, is shown as a circle with a radius of 1.2", which
corresponds to the X-ray signal extraction region.  No unresolved X-ray emission is observed at the BCG position. Lower 
panels:  the same as in the upper panels for SPT-CL J2344-4243 at $z=0.596$ (the Phoenix cluster), which shows a prominent 
unresolved emission in the hard band.}
\end{center}
\end{figure}

In the redshift range $0.2<z<0.3$, we have 40 clusters with HST images.  Seven clusters show a secondary BCG that may 
be associate with a minor or comparable mass halo that recently merged with the main cluster.  We find a secondary BCG in A2163 
\citep[see][]{2008Maurogordato}, AS0592, RXC J1514.9-1523 \citep{2015Kale}, A1682 \citep{2013Macario},
Z5247 \citep{2015Kale}, CL 2341.1+0000 and 1E0657-56 \citep{2006Clowe}, most of which are well-known major mergers 
where the merging halos can be clearly identified in the X-ray image.  In one case \citep[A2465, see][]{2011Wegner} we 
identify two separate X-ray halos belonging to the same superstructure.  Therefore we finally consider 81 BCGs out of 73 
clusters and groups.  The BCG list in the redshift bin  $0.2<z<0.3$, with redshift, position, and relevant references, is 
shown in Table \ref{clusterlist_lowz}.

In the redshift range  $0.55<z<0.75$, we have only 20 clusters with HST images.  Only 1 cluster is reported to have
3 BCGs \citep[MACS J0025.4-1222,][]{2008Bradac}.  Therefore, we finally consider 51 BCGs out of 49 clusters 
and groups.   In the high-redshift sample, some positions are based uniquely on the X-ray centroid (8 cases out of 51), since 
we are not able  to find the identification of the BCG and its position in the literature, nor to do this on the basis of available 
optical data. However, in all these cases, there are no hints of unresolved X-ray emission
from a BCG embedded in the ICM, so this choice does not affect our results as far as the X-ray selection 
function is concerned. The BCG list in the redshift bin  $0.55<z<0.75$, with redshift, BCG position and relevant
references, is shown in Table \ref{clusterlist_highz}. 

As an example, we show in Figure \ref{HST_images} two different cases: MS 0735.6+7421 at $z=0.216$ (upper panels) and 
SPT-CL J2344-4243 at $z=0.596$ (the Phoenix cluster, lower panels).  In both cases the position of the BCG is chosen from the 
HST image.  For MS 0735.6+7421 the HST image, on the left, is taken with ACS
with the F850LP filter (PI: McNamara).  The hard-band {\sl Chandra} image, on the right, shows no 
unresolved emission at the center, although MS 0735.6+7421 is one of the most powerful outbursts 
known to date \citep{2005McNamara,2007Gitti}. The ICM X-ray emission within the extraction radius, shown as
a circle, is used to set the upper limit to a possible sub-threshold unresolved emission in the hard band. 
In the second case, we use the HST image of the well-known Phoenix cluster \citep{2013bMcDonald}, taken with WFC3
with the F814W filter (PI: M. McDonald).  In this case, the hard-band {\sl Chandra} image shows strong
unresolved emission that dramatically overwhelms the surrounding ICM.  The challenge here is to establish well-defined
criteria for photometry to treat the many intermediate cases between these two extreme examples.

\begin{landscape}
\begin{deluxetable}{ccccc} 
\tablewidth{0pt}
\tablecaption{List of BCGs in X-Ray Groups and Clusters Observed by {\sl Chandra} in the Redshift Range $0.2<z<0.3$ with 
a Total Exposure of More than 20 ks. The redshift is listed in column 2, while the optical position of the BCG is listed in columns 
3 and 4.  The dataset used to measure the BCG position is listed in column 5, together with the corresponding reference when 
available. ``HST", with no reference, means that the positon of the BCG nucleus has been obtained directly from HST images. 
If a reference is listed 
first, followed by the name of the counterpart, the position is taken from the literature.  In the other cases, we obtain the position from NED (the counterpart name is also listed) }
\tablehead{  \colhead{Cluster} & \colhead{$z$}   &  \colhead{ $R.A._{BCG}$ }  & \colhead{ $Decl._{BCG}$ } & \colhead{ References } }
\startdata
G257.34-22.18	 & 0.2026	 & 06:37:14.5 & -48:28:23  &  NED, 2MASX J06371455-4828214  \\ 
CL 1829.3+6912	 & 0.2030	 & 18:29:05.7 &  +69:14:06 &   NED,  2MASX J18290571+6914064 \citet{2012Murgia} \\ 
A2163	BCG-1			& 	0.2030	 & 	   16:15:48.9 &   -06:08:41     &  HST, \citet{2008Maurogordato}  \\ %
A2163	BCG-2			& 	0.2030	 & 	  16:15:33.5   & -06:09:16      &  HST,  \citet{2008Maurogordato}   \\ 
A963	 				& 	0.2060	 & 	    	10:17:03.65	 & +39:02:49.6	& HST, \citet{2009Coziol}  \\    
RX J0439-0520	 & 0.2080	 &  04:39:02.23  &   +05:20:44    &  \citet{2015Kale},  2MASX J04390223+0520443  \\ 
G286.58-31.25	 & 	0.2100	 & 	  5:31:30.240 & -75:11:02.40 & \citet{2016Rossetti}   \\ 
RX J1256.0+2556	 & 	0.2120	 &      12:56:02.30 &  +25:56:36.51  &    HST   \\ 
ZW 2701	 & 	0.2140	 & 	  9:52:49.066 & +51:53:06.5      &  HST,  \citet{2015Kale}, 2MASX J09524915+5153053 \\    
RXC J1504-0248	 & 	0.2153	 & 	  15:04:07.49 &  -02:48:17.45   &  HST, \citet{2015Kale}  \\ 
MS 0735.6+7421	 & 	0.2160	 & 	     07:41:44.55 & +74:14:37.9     &   HST   \\   
A773	 & 	0.2170	 & 	   9:17:53.416 & +51:43:36.95    &  HST,  \citet{2015Kale},  2MASX J09175344+5143379\\   
G256.55-65.69 & 0.2195 &  02:25:53.16 & -41:54:52.49  &  NED, LCRS B022355.8-420821 \citet{2016Rossetti}\\
RXC J0510.7-0801 & 0.2195  & 05:10:47.9 & -08:01:45.00 & \citet{2015Kale}, 2MASX J05104786-0801449\\
MS 1006.0+1202	& 0.2210 &  10:08:47.74 &  +11:47:38.7     &  NED,  2MASX J10084771+1147379 \\
AS0592	BCG-1	 & 	0.2216	 & 	 6:38:48.605  & -53:58:24.37     &  HST \\ 
AS0592	BCG-2	 & 	0.2216	 & 	 6:38:45.15  & -53:58:22.11     &   HST \\ 
RXC J1514.9-1523	BCG-1	 & 0.2226	 &  15:14:57.59 &  -15:23:43.39 &  \citet{2015Kale}, 2MASX J15145772-1523447  \\ 
RXC J1514.9-1523	BCG-2	 & 0.2226	  &  15:15:03.1 & -15:21:53.0  &   \citet{2015Kale}, 2MASX J15150305-1521537 \\ 
A1763 		& 	0.2230	 &	13:35:20.11  &   +41:00:03.4  & HST \\          	
PKS 1353-341	 & 	0.2230	 & 		13:56:05.46 	&  -34:21:10.94	        & NED, \citet{2003Fomalont} \\ 
A1942	 & 	0.2240	 & 	   14:38:21.84 &  +03:40:13.05   &  NED, SDSS J143821.32+034013.4  \\ 
A2261	 & 	0.2240	 & 	 	17:22:27.20	 & +32:07:57.1  	&  HST, \citet{2009Coziol}   \\ 
1RXS J060313.4+421231	 & 0.2250	 & 06:03:16.7 & +42:14:41    &  HST, NED, 2MASX J06031667+4214416, \citet{2012vanWeeren}   \\ 
A2219	 & 	0.2256	 & 	 16:40:19.8 &  +46:42:41    &  HST, \citet{2015Kale}, 2MASX J16401981+4642409  \\ 
CL 0823.2+0425	 & 	0.2256	 &    08:25:57.8 & +04:14:48.0     & NED, 2MASX J08255782+0414480   \\ 
CL 0107+31	 & 	0.2270	 & 	 01:02:13.5     &  +31:49:24       &  NED,  2MASX J01021352+3149243  \\ 
A2390	 & 	0.2280	 & 	 21:53:36.82  & +17:41:43.86     &  HST, \citet{2015Kale}   \\ 
A2111	& 0.2290 &  15:39:40.5  &  +34:25:27  &   NED, \citet{2015Kale}, 2MASX J15394049+3425276  \\ 
A2667	 & 	0.2300	 & 		23:51:39.40	  &    -26:05:03.3  	&  HST, \citet{2009Coziol}   \\ 
RX J0439.0+0715	 & 	0.2300	 & 	  04:39:00.5  & +07:16:04     &  HST, \citet{2015Kale}, 2MASX J04390053+0716038  \\ 
RX J0720.8+7109	 & 	0.2309	 &      07:20:53.9    &    +71:08:59.4    &   HST, NED, 2MASX J07205404+7108586 \citet{2006Mulchaey}        \\ 
A267	 & 	0.2310	 & 	  	01:52:41.98	 & +01:00:26.4       & HST, \citet{2009Coziol}  \\ 
G342.31-34.90	 & 	0.2320	 &  20:23:20.0 & -55:36:03    &  NED,   2MASX J20232005-5536035  \\ 
A746	 & 	0.23225	 & 	  09:09:18.46 & +51:31:27.98      & NED,   2MASX J09091846+5131271 \\ 
A1682	 BCG-1 & 	0.2339	 & 	 13:06:50.1     & +46:33:33.1    &  HST   \\ 
A1682	 BCG-2 & 	0.2339	 & 	 13:06:45.731    & +46:33:30.12   &  HST     \\ 
A2146	 & 	0.2343	 & 	15:56:13.953 &  +66:20:53.62   &   HST, \citet{2015Kale}, 2MASX J15561395+6620530  \\ 
RXC J1459.4-1811	 & 	0.2357	 & 	14:59:28.75    &   -18:10:45.18  & \citet{2015Kale}, 2MASX J14592875-1810453   \\  
G347.18-27.35	 & 	0.2371	 & 19:34:54.4 &-50:52:19.2  &  NED,  [GSB2009] J193454.46-505218.5 \\ 
G264.41+19.48	 & 0.2400	 &  10:00:01.4    &   -30:16:33.0  &  NED, 2MASX J10000143-3016331   \\ 
4C+55.16	 & 	0.2411	 & 8:34:54.900    &  +55:34:21.11   &   HST, NED WHL J083454.9+553421 \citet{2009Wen}\\ 
Z5247	BCG-1 &0.24305	 & 12:34:24.100 & +09:47:16.00  & NED, \citet{2015Kale}, 2MASX J12342409+0947157 \\ 
Z5247	BCG-2 &0.24305	 &   12:34:17.567  &   +9:45:58.16  &   HST \\ 
A2465-1	 & 	0.2453	 & 22:39:39.6 & -05:43:56     &  NED, [W2011] J339.91522-05.73214   \citet{2011Wegner} \\
A2465-2	 & 0.2453	 &    22:39:24.6 & -05:47:17    &  NED, 2MASX J22392454-0547173  \citet{2011Wegner} \\ 
A2125	 & 	0.2465	 &   15:41:01.98   &   +66:16:26.56   &   HST \\ 
CL 2089	 &	0.2492	 & 	  9:00:36.846    &  +20:53:40.14   &   HST, \citet{2015Kale} 2MASX J09003684+2053402  \\ 
RX J2129.6+0005	 & 	39.96       &    21:29:39.952   &  +00:05:21.15   &  HST, 2MASX J21293995+0005207  \\ 
RCS0222+0144	 & 	23.33       &   02:22:40.9 	& +01:44:42    & NED, RCS 01200101288     \\ 
A2645	 & 	0.2510	 & 	23:41:17.022   &   -09:01:11.74 &  \citet{2015Kale} 2MASX J23411705-0901110   \\ 
A1835	 & 	0.2532	 & 	14:01:02.1 &   +02:52:42.5    &  HST, \citet{2015Kale} 2MASX J14010204+0252423  \\ 
A521	 & 	0.2533	 &  4:54:06.870   &   -10:13:24.79   & HST,  \citet{2015Kale} 2MASX J04540687-1013247  \\ 
RXC J1023.8-2715	 & 	0.2533	 &  10:23:50.21	   &   -27:15:23.99  &   NED, 2MASX J10235019-2715232  \\  
CL 0348	 & 	0.2537	 & 	 1:06:49.40    &   +01:03:22.66	   &   HST \\ 
MS 1455.0+2232	& 	0.2578  &   14:57:15.04 &   +22:20:33.60  &  HST  \\ 
G337.09-25.97	& 0.2600	& 19:14:37.3  & -59:28:20 & NED, [CB2012] J288.655540-59.472132 \citet{2012Chon}\\
SL J1204.4-0351	 & 0.2610	 &  12:04:24.3   & -03:51:10  &   NED, 2MASX J12042431-0351096 \\ 
G171.94-40.65 & 	0.2700	 & 	 3:12:57.499 &  +8:22:10.88     &   HST   \\ 
A2631	 & 	0.2730	 &   23:37:39.76	& +0:16:17.0    & HST,  \citet{2009Coziol}  \\ 
G294.66-37.02	 & 	0.2742	 &  3:03:46.224  &-77:52:43.32    &   \citet{2016Rossetti}      \\ 
G241.74-30.88	 & 0.2747	 & 	 05:32:55.6 & -37:01:36      & NED, [GSB2009] J053255.66-370136.1 \\ 
RXC J2011.3-572 & 0.2786 & 20:11:26.9 & -57:25:11 & NED, SPT-CL J2011-5725 BCG \citet{2009Guzzo}\\ 
A1758	 & 	0.2790	 &  13:32:38.4  & +50:33:35   &  HST, \citet{2015Kale}  2MASX J13323845+5033351   \\ 
G114.33+64.87	 & 	0.2810	  & 13:15:05.2      &    +51:49:03	 &  HST    \\ 
A697	 & 	0.2820	 & 	 8:42:57.63   &  +36:22:01    &  HST  \\ 
CL 2341.1+0000  BCG-1	 & 	0.2826	 &  23:43:40.06 & +0:18:21.76    &   NED, SDSS J234340.07+001822.3 \\  
CL 2341.1+0000	BCG-2 & 	0.2826	 &  23:43:35.68 & +0:19:50.70       &   NED, SDSS J234335.66+001951.4    \\  
RXC J0232.2-4420	 & 	0.2836	 & 	 2:32:18.57   &   -44:20:48   &  HST    \\ 
RXC J0528.9-3927	 & 	0.2839	 &  05:28:52.99 & -39:28:18.1     & NED,  [GSB2009] J052852.99-392818.1  \citet{2009Guzzo} \\ 
A611	 & 	0.2880	 &  8:00:56.83    &   +36:03:23.5  &   HST  \\ 
3C438	 & 	0.2900	 &    	21:55:52.25   &  +38:00:28.35   &    HST \\ 
ZW 3146	 & 	0.2906	 & 	  10:23:39.7   &   +4:11:10.7  & HST, \citet{2015Kale} 2MASX J10233960+0411116  \\ 
G195.62+44.05	 & 	0.29165	 & 	  9:20:25.756	& +30:29:37.74     & HST   \\ 
RX J0437.1+0043	 & 	0.2937	 &	 04:37:09.5 &   +0:43:51   &   \citet{2015Kale}, 2MASX J04370955+0043533  \\  
A2537	 & 	0.2950	 & 	 23:08:22.3  &  -02:11:33.2  & HST,  \citet{2015Kale}, 2MASX J23082221-0211315 \\ 
G262.25-35.36	& 0.2952 &  05:16:37.2 &  -54:30:59 &  NED, SSTSL2 J051637.18-543059.3 \citet{2009Coziol} \citet{2016Rossetti} \\ 
1E0657-56	 BCG-1 & 	0.2960	 & 	6:58:38.073  & -55:57:26.06      &    HST \\ 
1E0657-56	 BCG-2  & 	0.2960	 &  6:58:16.089    &	 -55:56:35.33     &   HST  \\ 
Abell S295 	 & 		0.3		& 		2:45:24.812	 & -53:01:45.56   	&  HST	\\
G292.51+21.98 	 	& 	0.3  & 		    12:01:04.953    & -39:51:55.14 & \citet{2016Rossetti} \\  
\enddata
\label{clusterlist_lowz}
\end{deluxetable}
\end{landscape}

\begin{landscape}
\begin{deluxetable}{ccccc} 
\tablewidth{0pt}
\tablecaption{List of BCGs in X-Ray Groups and Clusters Observed by {\sl Chandra} in the Redshift Range $0.55<z<0.75$ with 
a Total Exposure of More than 20 ks. The redshift is listed in column 2, while the optical position of the BCG is listed in 
columns 3 and 4.  The dataset used to measure the BCG position is listed in column 5, together with the corresponding reference 
when available.  ``HST" means that the positon of the BCG nucleus has been obtained from HST images and it is found
consistent with the corresponding reference.  If a reference is 
listed first, followed by the name of the counterpart, the position is taken from the literature.  In the other cases, we obtain the 
position from NED (the counterpart  name is also listed).  Finally, in eight cases the position is taken directly from the X-ray surface 
brightness peak (``X-ray")}
\tablehead{  \colhead{Cluster} & \colhead{$z$} & \colhead{ $R.A._{BCG}$ }  & \colhead{ $Decl._{BCG}$ } & \colhead{ References } }
\startdata
ACT J0346-5438  & 0.55 &    	03:46:55.5	 & -54:38:55	&   NED, ACT-CL J0346-5438 BCG \\ 
MS 0451.6-0305  & 0.55 	&  	4:54:10.905   & -3:00:52.41    	&   HST, \citet{2010Berciano}  \\		
V1121+2327	 & 	0.562 & 	   11:20:56.77   &        +23:26:27.87   &    NED,  \citet{2011Szabo}  \\  
CL 1357+6232	 & 	0.5628 & 	  13:57:16.8   & +62:32:49.6         &  NED, \citet{2011Szabo}        \\ 
SPT-CL 2332-5051	 & 	0.5707	 & 	  23:31:51.123	    &      -50:51:53.94       &   HST, \citet{2013bMcDonald}      \\ 
SPT-CL J2148-6116	 & 	0.571	 & 	 21:48:42.720	 &  -61:16:46.20   & NED, \citet{2016McDonalda}   \\  
CL 0216-1747	 & 	0.578	 &       	2:16:32.632    &     -17:47:33.17       &    HST, \citet{2002Perlman}   \\ 
CL 0521-2530	 & 0.581	 & 	   05:21:10.5   &  -25:31:06.5   &  X-ray,  \citet{2007Burenin,2012Mehrtens}  \\  
MS 2053.7-0449	 & 	0.583	     &   20:56:21.47      &    -4:37:50.1        &  HST, \citet{2007Verdugo}      \\ 
MACS 0025.4-1222  BCG1 & 0.584  & 	 0:25:33.018  & -12:23:16.80   & HST, \citet{2008Bradac}  \\
MACS 0025.4-1222  BCG2 & 0.584  &   	0:25:32.021  & -12:23:03.80 		 & HST, \citet{2008Bradac}  \\
MACS 0025.4-1222  BCG3 & 0.584  & 	  0:25:27.380  &  -12:22:23.00   		 & HST, \citet{2008Bradac}  \\
SDSS J1029+2623	 & 	0.584	 & 10:29:12.456  &	+26:23:31.91     &   HST, \citet{2012Ota}    \\ 
CL 0956+4107	 & 	0.587 &  09:56:02.874 & +41:07:20.33       &     NED, \citet{2011Szabo}     \\  
MACS 2129.4-0741	 & 	0.5889	 &    21:29:26.056  &  -7:41:28.95   &   \citet{2010Stern}     \\  
ACT J0232-5257	 & 	0.59	 & 	   	02:32:42.80 & -52:57:22.3      &      NED, \citet{2013Sifon}    \\  
CL 0328-2140	 & 	0.59 &    03:28:13.6  &      -21:40:19     &    NED, \citet{2015Liu}  \\  
MACS 0647.7+7015	 & 	0.5907	  & 6:47:50.23 & +70:14:54.01        &  HST, \citet{2010Stern}      \\ 
RX J1205	 & 	0.5915	 &  12:05:51.372	&  +44:29:09.30        & HST,   \citet{2007Jeltema}  \\ 
SPT-CLJ 2344-4243	 (Phoenix) & 0.596	    &     23:44:43.95  &  -42:43:12.86   &   HST, \citet{2012McDonald}     \\ 
CL 1120+4318	 & 	0.60	 & 	   11:20:07.4 &  +43:18:07       & X-ray, \citet{2007Burenin}       \\  
ACT J0559-5249	 & 	0.6112	 &     	5:59:41.644 &      -52:50:02.39       &    HST, \citet{2013Sifon}    \\ 
CL 1334+5031	 & 	0.62	 & 	  	13:34:20.563        &   +50:31:03.91        &   HST,  \citet{2011Adelman}    \\ 
RCS 1419+5326	 & 	0.62	 & 		14:19:12.148	& +53:26:11.47  & HST, \citet{2013Ebeling}  \\
SPT-CL J0417-4748	 & 	0.62	 & 	   4:17:23.0      &     -47:48:45.6       &  NED, \citet{2013bMcDonald}   \\ 
SPT-CL J0256-5617	 & 	0.63	 & 	   	2:56:25.344 & -56:17:52.08   &      X-ray, \citet{2013Reichardt}   \\ 
SPT-CL J0426-5455	 & 0.63 &  04:26:04.1   &  -54:55:31        &  NED, \citet{2013Reichardt}    \\  
CL J0542.8-4100	 & 	0.64	 & 	   05:42:50.1 & -41:00:00         &  X-ray, \citet{2013bMcDonald}   \\ 
SPT-CL J0243-5930	 & 	0.65	 & 02:43:27.0   &      -59:31:01.88     &    NED, \citet{2012Song}    \\ 
SPT-CL J0352-5647	 & 	0.66	 &     03:52:56.8   &      -56:47:57      &        NED, \citet{2012Song}    \\ 
LCDCS 954	 & 	0.67	 & 	    	14:20:29.7    &   -11:34:04       &   NED, \citet{2001Gonzalez}   \\      
ACT-CL 0206-0114	 & 	0.676	  &  	02:06:22.79      &     -01:18:32.5       &  NED, \citet{2013Wen}    \\    
CL 1202+5751	 & 	0.677	 &      	12:02:13.7  &     +57:51:53      &   X-ray, \citet{2007Burenin}    \\  
DLS J1055-0503	 & 	0.68	 & 	   10:55:12.0     &       -05:03:43      &   NED, \citet{2006Wittman}     \\   
SDSS J1004+4112	 & 	0.68	 & 	  10:04:34.18	    &     +41:12:43.57         &  HST,  \citet{2012Oguri}     \\ 
CL 0405-4100	 & 	0.686 & 	    	04:05:24.3  & -41:00:15      &   X-ray, \citet{2007Burenin}   \\   
RX J1757.3+6631	 & 	0.691	 & 	 17:57:19.6       &    +66:31:33       &     NED, \citet{2013Rumbaugh}  \\   
MACS 0744.8+3927	 & 	0.6976	 & 	  	7:44:52.770 &  +39:27:25.55       &    HST, \citet{2011Zitrin}    \\ 
RCS 2327-0204	 & 	0.70	 & 	 23:27:27.6     &    -02:04:37         & HST,  \citet{2012Rawle} \\ 
SPT-CL 0528-5300	 & 	0.70	 & 	   05:28:05.3     &     -52:59:53    &   NED, \citet{2010Menanteau}   \\  
V1221+4918	 & 	0.70	 & 		12:21:24.5 &  +49:18:13         & X-ray, \citet{1998Vikhlinin}      \\   
ACT J0616-5227	 & 	0.71	 & 	06:16:34.2      &     -52:27:13     &   NED, \citet{2010Menanteau}     \\   
SDSS J022830.25+003027.9	 & 	0.72	       &   02:28:25.9   &      +00:32:02     &     NED, \citet{2010eWen}   \\  
CL J2302.8+0844	 & 	0.722	 & 	23:02:48.1      &     +08:43:51       &    radio, \citet{1998Condon}  \\  
SPT-CL J2043-5035	 & 	0.723	 & 	 	20:43:17.53	 & -50:35:32.4    &   HST, \citet{2012Song}     \\ 
CL J1113.1-2615	 & 	0.725	 & 	  	11:13:05.2   &      -26:15:39      &   X-ray, \citet{2010Evans}     \\ 
RCS 1107.3-0523	 & 0.735 &	11:07:24.066  &   -05:23:20.83    &    NED, \citet{2014Bai}    \\ 
3C254	 & 	0.736619	 & 	11:14:38.747     &     +40:37:20.56      &  HST, \citet{2010Evans}     \\ 
SPT-CL 0001-5748 & 0.74 & 0:01:00.033  &  -57:48:33.42	 & HST, \citet{2012Song}	\\
SPT-CL 0324-6236	 & 	0.74	 &   03:24:12.2    &      -62:35:56       &   HST, \citet{2012Song}      \\ 
ACT J0102-4915  & 0.75  &		 1:02:57.844   &	-49:16:19.14 		& HST, \citet{2013Hilton}  \\
\enddata
\label{clusterlist_highz}
\end{deluxetable}
\end{landscape}

\section{Data Reduction and Analysis}

\subsection{Data Reduction}
\noindent
The lists of the {\sl Chandra} data used for each cluster, with total {\sl Chandra} exposure time and observing mode, are shown in 
Table \ref{obsid_lowz} and \ref{obsid_highz} for the low- and high- redshift bin, respectively. We performed a standard data 
reduction starting from the level=1 event files, using the {\tt CIAO 4.9} software package, with the most recent version of the 
{\sl Chandra} Calibration Database ({\tt CALDB 4.7.3}). When observations are taken in the VFAINT mode, we ran the task {\tt
acis$\_$process$\_$events} to flag background events that are most likely associated with cosmic rays and removed them.  With 
this procedure, the ACIS particle background can be significantly reduced compared to the standard grade selection.  The
data are filtered to include only the standard event grades 0, 2, 3, 4, and 6.  We checked visually for hot columns left after the 
standard reduction.  For exposures taken in VFAINT mode (the large majority of our dataset), there are practically
no hot columns or flickering pixels left after filtering out bad events.
We also applied CTI correction to ACIS-I data.  We finally filtered time intervals with high background by performing a
$3\sigma$ clipping of the background level using the script {\tt analyze\_ltcrv}.  The final effective exposure times are generally 
very close to the original observing time.  Our data reduction is not affected by possible undetected flares or other background 
related issues, since the background at the BCG position is swamped by the surrounding ICM emission.  
The fully reduced data (event 2 files) are used to create the soft-band (0.5-2 keV) and hard-band (2-7 keV) images.  
The choice of a relatively narrow hard band is justified by the necessity of minimizing the background while
leaving the bulk of the source signal in the image.  The use of the 2-7 keV band in this respect is based on our previous 
experience in detecting faint sources in X-ray deep fields \citep[see][]{2002RosatiCDFS}.
We also produce soft- and hard-band combined exposure maps (in cm$^2$) computed at the monochromatic energies
of 1.5 and 4.0 keV, respectively.  The exposure maps are used to compute the small correction for vignetting
in our aperture photometry of the BCG, and the more significant correction for the cool-core strength parameter.

\begin{deluxetable}{cccc} 
\tablewidth{0pt}
\tablecaption{ {\sl Chandra} Data Used in This Work for Clusters in the Redshift Range $0.2<z<0.3$.
The total exposure time in ks after data reduction is listed in Column 2.  
}
\tablehead{  \colhead{Cluster}   &  \colhead{Exptime  (ks)} & \colhead{ObsIDs} & \colhead {Detector and 
Observing Mode} }
\startdata
G257.34-22.18	 &  	24.65   & 15125  &  ACIS-I, VFAINT \\
CL 1829.3+6912	 & 	64.60  & 10412, 10931   & ACIS-I, VFAINT  \\ 
A2163			         & 	80.43        & 1653, 545 &  ACIS-I, VFAINT  \\ 
A963	 				 & 	36.19       &  903  & ACIS-S, FAINT \\    
RX J0439-0520	 	 & 28.42  		& 9369, 9761  & ACIS-I, VFAINT\\
G286.58-31.25	 & 	22.16       & 15115 & ACIS-I, VFAINT \\ 
RX J1256.0+2556	 & 	25.37       & 3212  &    ACIS-S, VFAINT \\ 
ZW 2701	 		 	& 	121.90      & 3195, 12903  & ACSI-S, VFAINT\\    
RXC J1504-0248	& 	148.13      &    5793, 17197, 17669, 17670 &   ACIS-I, VFAINT\\ 
MS 0735.6+7421	& 	474.62     & 10470, 10468, 10469 &    ACIS-I, VFAINT \\   
	 						& 	       			& 10471, 10822, 10918, 10922  &  \\   
A773	              	& 	40.43       & 533, 3588, 5006  & ACIS-I, VFAINT \\   
G256.55-65.69 	& 28.67 & 17476, 15110 & ACIS-I, VFAINT \\
RXC J0510.7-0801 & 20.70 &  14011 & ACIS-I, VFAINT \\
MS 1006.0+1202	 & 67.58 &  925, 13390 &  ACIS-I, VFAINT\\
AS0592		 		& 	107.69      &  9420, 15176, 16572, 16598  &  ACIS-I, VFAINT\\ 
RXC J1514.9-1523 & 	50.71    & 15175 & ACIS-I, VFAINT \\ 
A1763 				 & 19.50 & 3591	&  ACIS-I, VFAINT \\          	
PKS 1353-341	 	 & 	30.25       & 17214 	& ACIS-I, VFAINT\\ 
A1942	 			& 	61.40   & 7707, 3290  & ACIS-I, VFAINT\\ 
A2261	 			 & 	33.39       &  550, 5007 	&  ACIS-I, VFAINT \\ 
1RXS J060313.4+421231	  & 	235.93       & 15171, 15172, 15323 & ACIS-I, VFAINT  \\ 
A2219	 			 & 	146.65       & 13988, 14355, 14356  & ACIS-I, VFAINT  \\ 
	 			 		& 			       & 				14431, 14451  &   \\ 
CL 0823.2+0425		 & 21.22  & 10441 &  ACIS-I, VFAINT  \\ 
CL 0107+31	 		 & 	48.25       & 521 &  ACIS-I, FAINT\\ 
A2390	 	 	& 	92.89       &  4193 &   ACIS-S, VFAINT\\ 
A2111			 & 20.88 & 11726  &  ACIS-I, VFAINT\\ 
A2667	 		 & 9.65       &   2214 &  ACIS-S, VFAINT\\
RX J0439.0+0715	  & 	19.02       & 3583    &  ACIS-I, FAINT \\ 
RX J0720.8+7109	 	 & 	117.26      &  13984, 14449, 14450 &   ACIS-S, VFAINT  \\ 
A267	 			 & 	19.88      &  3580  & ACIS-I, VFAINT\\ 
G342.31-34.90	 & 20.81      &   15108 & ACIS-I, VFAINT \\ 
A746		 		& 	25.73       &15191  & ACIS-I, VFAINT \\ 
A1682			 	& 	29.55       &  3244, 11725 &ACIS-I, VFAINT  \\ 
A2146	 			& 	375.34   &  1224, 12246, 12245 & ACIS-I, VFAINT\\
 						& & 13020, 13021, 13023& \\
						& & 13120, 13138 & \\ 
RXC J1459.4-1811	&  39.63  & 9428	&  ACIS-S, VFAINT\\  
G347.18-27.35	& 24.66 & 15120 & ACIS-I, VFAINT \\ 
G264.41+19.48	 & 30.58   &15132  &  ACIS-I, VFAINT \\ 
4C+55.16	 		& 	89.86     & 4940  	& ACIS-S, VFAINT \\ 
Z5247	 				& 29.66  & 539, 11727& ACIS-I, VFAINT \\ 
A2465 					 & 	69.15   &14010,  15547 &  ACIS-I, VFAINT\\
A2125	 				& 	86.03       &  2207, 7708 & ACIS-I, VFAINT\\ 
CL 2089	 			 & 	40.64  & 10463 &    ACIS-S, VFAINT\\ 
RX J2129.6+0005	& 	39.52       & 552, 9370   & ACIS-I, VFAINT \\ 
RCS 0222+0144	& 	23.24       & 10485   &  ACIS-S, VFAINT  \\
A2645					& 	23.46       &  14013  &  ACIS-I, VFAINT \\ 
A1835	 				 & 	193.20       &  6880, 6881, 7370  &    ACIS-I, VFAINT \\ 
A521					& 	127.03      &  901, 12880&  ACSI-I, VFAINT\\ 
RXC J1023.8-2715 & 	36.38      &  9400 & ACIS-S, VFAINT\\  
CL 0348	 			 & 	48.73       & 10465 &   ACIS-S, VFAINT\\ 
MS 1455.0+2232	 & 	98.85 &  4192, 7709  & ACIS-I, VFAINT\\ 
G337.09-25.97	 & 	24.75 &  15135 & ACIS-I, VFAINT\\
SL J1204.4-0351	 & 22.64 &  12304  & ACIS-I, VFAINT\\ 
G171.94-40.65	 & 	26.63       &  15302 &   ACIS-I, VFAINT \\ 
A2631	 				& 	25.99       &  11728, 3248 & ACIS-I, VFAINT\\ 
G294.66-37.02	& 	33.64       & 15113 &   ACIS-I, VFAINT \\ 
G241.74-30.88	 & 	24.75  & 15112  & ACIS-I, VFAINT \\ 
RXC J2011.3-572  &   23.90  & 4995  &  ACIS-I, VFAINT\\ 
A1758	 				& 	153.97    & 15538, 15540, 13997, 7710  &   ACIS-I, VFAINT \\ 
G114.33+64.87	 & 	77.18       & 16126, 15123&  ACIS-I, VFAINT \\ 
A697					& 	19.49      & 4217 &  ACIS-I, VFAINT\\ 
CL 2341.1+0000	 & 	222.74      & 17490, 17170, 18702, &     ACIS-I, VFAINT  \\  
							 & 					    &  18703, 5786 &       \\  
RXC J0232.2-4420	 & 	22.51       & 4993 &    ACIS-I, VFAINT \\ 
RXC J0528.9-3927		 & 	105.63       & 15658, 15177, 4994 &   ACIS-I, VFAINT \\ 
A611					 & 	35.72       &   3194&  ACIS-S, VFAINT \\ 
3C438	 				 & 	158.31       &  12879, 13218, 3967 &   ACIS-S, VFAINT \\ 
ZW 3146	 			& 	39.87  &    9371 & ACIS-I, VFAINT  \\ 
G195.62+44.05		 &  45.06	&  15128, 534 &  ACIS-I, VFAINT \\ 
RX J0437.1+0043	  &	42.54  &  11729, 7900 & ACIS-I, VFAINT \ \\  
A2537	 				 & 	38.41       &  9372 &  ACIS-I, VFAINT \\ 
G262.25-35.36	& 30.70  & 15099, 9331 &   ACIS-I, VFAINT \\ 
1E0657-56	 		 & 544.76       & 5361, 5358, 5357 & ACIS-I, VFAINT \\ 
					 		 & 			       & 5356, 5355, 4986			 & 			 \\ 
							 &       				& 4985, 4984, 3184  &  \\ 
Abell S295 	 		& 	205.23      & 16526, 16525, 16524, 16127  	& 	ACIS-I, VFAINT \\
							& 	     			 &  16282, 12260  	& 	 \\
G292.51+21.98 	 & 		42.68    & 	15134  & ACIS-I, VFAINT	\\ 
\enddata
\label{obsid_lowz}

\end{deluxetable}

\begin{deluxetable}{cccc} 
\tablewidth{0pt}
\tablecaption{ {\sl Chandra} Data Used in This Work for Clusters in the Redshift Range  $0.55<z<0.75$.   
The total exposure time in ks after data reduction is listed in Column 2. }
\tablehead{  \colhead{Cluster}   &  \colhead{Exptime  (ks)} & \colhead{ObsIDs} & \colhead {Detector and 
Observing Mode} }
\startdata
ACT J0346-5438  			&     34.05     &    12270, 13155     &   ACIS-I, VFAINT      \\
MS 0451.6-0305  		&  	42.41         &      902   &   ACIS-S, FAINT      \\
V1121+2327	 			 & 	   70.05      &     1660    &    ACIS-I, VFAINT        \\
CL 1357+6232				& 	  43.76            &   5763, 7267      &    ACIS-I, VFAINT       \\
SPT-CL 2332-5051	 	 & 	34.51          &    9333, 11738    &      ACIS-I, VFAINT     \\
SPT-CL J2148-6116	 	 &   36.10	       &    13488     &      ACIS-I, VFAINT         \\
CL0216-1747	 			 &     61.84       &      5760, 6393   &      ACIS-I, VFAINT         \\
CL0521-2530	 			 & 	  33.69        &  5758,  6173, 4928     &      ACIS-I, VFAINT      \\
MS 2053.7-0449	 	     &     44.30       &    1667     &        ACIS-I, VFAINT     \\
MACS 0025.4-1222  	 & 	   157.02      &   10413, 10797, 10786     &     ACIS-I, VFAINT     \\
								 	 & 	   				     &   		5010,3251      &          \\
SDSS J1029+2623			 &     55.67     &     11755    &     ACIS-S, VFAINT    \\
CL 0956+4107	 			&      59.20    &     5759, 5294, 4930    &        ACIS-I, VFAINT   \\
MACS 2129.4-0741	 	&    36.67         &    3199, 3595    &        ACIS-I, VFAINT     \\
ACT J0232-5257	 		 & 	19.69            &     12263    &       ACIS-I, VFAINT     \\
CL 0328-2140	 			 &       56.19     &    5755, 6258     &    ACIS-I, VFAINT     \\
MACS 0647.7+7015	 	  &    38.64      &   3196, 3584      &     ACIS-I, VFAINT     \\
RX J1205	 					 &    29.69       &    4162     &     ACIS-S, VFAINT    \\
SPT-CL J2344-4243	 		&   129.07          &   16545, 16135, 13401      &      ACIS-I, VFAINT      \\
CL 1120+4318	 			& 	19.74         &   5771      &     ACIS-I, VFAINT       \\
ACT J0559-5249	 		&    108.15 	         &    13117, 13116, 12264     &        ACIS-I, VFAINT    \\
CL 1334+5031	 			 & 	 19.49 	         &    5772     &         ACIS-I, VFAINT    \\
RCS 1419+5326			 & 	56.27         &     5886, 3240    &      ACIS-S, VFAINT        \\
SPT-CL J0417-4748	 		 & 	21.78         &    13397     &    ACIS-I, VFAINT      \\
SPT-CL J0256-5617	 	 & 	25.63           &    14448, 13481     &    ACIS-I, VFAINT         \\
SPT-CL J0426-5455	 		&   32.23        &    13472     &      ACIS-I, VFAINT     \\
CL J0542.8-4100	 		 & 	 50.11           &   914      &      ACIS-I, FAINT    \\
SPT-CL J0243-5930			 &     46.94     &    13484,  15573    &      ACIS-I, VFAINT   \\
SPT-CL J0352-5647	 			 &     45.06         &    13490, 15571     &        ACIS-I, VFAINT    \\
LCDCS 954			 			& 	 28.56         &       5824    &     ACIS-S, VFAINT       \\
ACT J0206-0114	 		  &   29.69        &      16229   &     ACIS-I, VFAINT      \\
CL 1202+5751	 			&     58.39         &     5757    &    ACIS-I, VFAINT      \\
DLS J1055-0503	 		 & 	20.06          &   4212      &       ACIS-I, VFAINT  \\
SDSS J1004+4112	 			 & 	243.26         &    5794, 11546-11549  &    ACIS-S, VFAINT     \\
						 			 & 					         &    14495-14500  &         \\
CL 0405-4100				 & 	 76.70            &     7191, 5756    &    ACIS-I, VFAINT       \\
RX J1757.3+6631	 			 & 	46.45         &    10443, 11999     &   ACIS-I, VFAINT       \\
MACS 0744.8+3927	 		 & 	  	88.83	         &   6111, 3585, 3197     &    ACIS-I, VFAINT       \\
RCS 2327-0204				 & 	143.03         &     14361, 14025    &    ACIS-I, VFAINT       \\
SPT-CL 0528-5300	 		 & 	 123.84	         &   11874, 10862, 11747, 11996   &    ACIS-I, VFAINT    \\
								 		 & 	 	         &   12092, 13126, 9341   				&        							\\
V1221+4918	 				& 	78.39		         &       1662  &      ACIS-I, VFAINT     \\
ACT J0616-5227	 			& 		38.59         &   12261, 13127      &     ACIS-I, VFAINT    \\
SDSS J022830.25+003027.9	&   49.32       &      16303  &    ACIS-S, VFAINT        \\
CL J2302.8+0844	 			& 		107.97         &      918   &   ACIS-I, FAINT      \\
SPT-CL J2043-5035	 		 & 	78.99	         &    13478     &  ACIS-I, VFAINT          \\
CL J1113.1-2615	 			 & 	103.31     &     915    &     ACIS-I, FAINT      \\
RCS 1107.3-0523	 			&		93.71         &  5887,  5825      &     ACIS-S, VFAINT    \\
3C254	 							 & 		29.54         &   2209      &     ACIS-S, VFAINT      \\
SPT-CL 0001-5748			&	    30.14     &      9335   &    ACIS-I, VFAINT     \\
SPT-CL 0324-6236			 &   	54.83         &       12181, 13137, 13213    &      ACIS-I, VFAINT     \\
ACT J0102-4915 				 &			349.76         &  14022, 14023, 12258      &    ACIS-I, VFAINT      \\
\enddata
\label{obsid_highz}

\end{deluxetable}

\subsection{Detection of X-Ray Emission from the BCG}
\noindent
Only a small subset of groups and clusters host BCGs with an X-ray AGN, and it is a hard task to identify the associated
unresolved emission in the X-ray images.  In particular, we expect most of them to have moderate or low X-ray luminosity 
well below the ICM emission at the BCG position.  Therefore, the optical position is a crucial information 
to evaluate the X-ray emission or the corresponding upper limit for all the BCGs in our sample.
We stress that the measurement of the upper limits when no X-ray emission is visible is relevant
to firmly evaluate the actual flux detection limit of each image.  To identify and quantify the X-ray emission of the BCG, we 
select a circle of 1.2 arcsec radius at the position of the optical BCG, and an annulus with outer and inner radii of 3 and 1.5 
arcsec, respectively.  This choice is dictated mainly by the fact that at the {\sl Chandra} aimpoint, 
about 95\% and 90\% of the source emission is included 
in a circle with a radius of 1.2 arcsec at 1.5 and 4.0 keV, respectively.  In addition, we also need to evaluate the ICM 
emission as close as possible to the BCG.  Therefore, we limit the background estimation to a small annulus with a 
maximum radius of 3 arcsec.  This measurement is a good proxy of the background in the assumption that the ICM surface 
brightness is flat within 3 arcsec from the BCG position.  This choice is clearly an approximation, since the actual  
ICM emission at the BCG position is hard to estimate, especially in cool-core clusters. The ICM surface brightness can be 
enhanced with respect to the outer annulus due to a very peaked cool core or the presence of a compact corona 
\citep[see][]{2001Vikhlinin}, but, as often happens, it can also by significantly lower due to the presence of unnoticed 
cavities associated with the AGN radio-powered jets from the BCG itself.  On the other hand, cavities may also be present in 
the outer annulus, contributing additional uncertainties to the measurement of the ICM emission at the BCG position.  These 
uncertainties, due to the ubiquitous presence of cavities carved in the ICM, should be treated as a source of systematic error.  The 
robustness of our aperture photometry based on a constant surface brightness within the inner 3 arcsec is investigated in the 
next subsection, where we explore the background measurement on the basis of a more complex modeling of the surface 
brightness.

Under the assumption of flat surface brightness within 3 arcsec, the total background in the source region $B_S$ is obtained 
by geometrically scaling the number of counts observed in the outer annulus, therefore  $B_S\equiv 0.213\times B$, 
where $B$ is the total exposure-corrected  number of counts in the annulus, and $0.213$ is the fixed geometrical 
scaling factor.\footnote{Clearly, the presence of other unresolved sources in this region, would imply the removal of part of 
the annulus, and therefore a different scaling factor.  However, we found none.}  We define a source signal simply as 
$S=CTS_S-0.213\times B$, where $CTS_S$ is the total exposure-corrected number of counts found in the images in the 
inner 1.2 arcsec.   The source signal is computed in the soft (0.5-2 keV) and hard (2-7 keV) bands.  The statistical noise is 
computed as $N_{stat}\equiv \sqrt{CTS_S+B_S}$, and it does not include additional components associated with intrinsic 
fluctuations in the ICM surface brightness. 

In our approach, the direct photometry is a  model-independent but noisy estimator.  In particular, we should not rely 
on photometry alone to decide whether unresolved X-ray emission from the BCG is detected or not in our data.  Therefore 
we perform an accurate visual inspection to flag X-ray unresolved sources at the BCG position in one of the 
two bands.  Then, we consider the signal-to-noise ratio, $S/N_{stat}$, distribution measured for our sources in the soft and hard band in both 
redshift ranges, and select a S/N threshold appropriate for source detection.  This is important to compute also the 
actual detection limit of each image as well.   Finally, we do not attempt to refine or expand our search of unresolved 
emission with a spectral analysis, as proposed in \citet{2013Hlavacek-Larrondo}.  The main reason is that we wish to explore a 
large S/N range, therefore most of our sources, which have a low S/N, cannot be spectrally analyzed, and the hardness ratio is
too noisy to firmly identify the presence of nonthermal emisson.  Another reason is related to the possible presence of
a population of nonthermal electrons associated with mini-halos, which may contribute with some inverse Compton 
emission that might change the hardness ratio of the diffuse emission.  Therefore, all our conclusions on the presence of unresolved, 
nuclear emission in cluster cores is based on high-resolution photometry.  Eventually, 
only for the sources with clear unresolved emission can a detailed spectral analysis can be performed, as we
show in Section 4.4.

\subsection{Evaluation of Systematic Uncertainties in Aperture Photometry}
\noindent
The scale of 3 arcsec, within which we assume a flat  surface brightness profile, corresponds to a physical scale of 
10.0-13.5 kpc and 19.5-22.0 kpc for the low- and high- redshift sample, respectively.  The chemical and thermodynamical 
properties of the ICM can vary significantly on this scale, and such variations can create positive or negative fluctuations in surface 
brightness, in particular driven by turbulent motions (e.g., \citealt{2013bGaspari,2016Khatri}).  While on the one hand, the surface brightness 
is expected to increase following the typical behaviour of a cool core, the feedback activity of the BCG may instead produce 
cavities, reducing the ICM emission close to the BCG.  Moreover, bright and compact X-ray coronae may be still present in the 
center of BCGs \citep[see][]{2001Vikhlinin}, although such coronae have small kiloparcsec-scale size, with extension below the resolution 
limits.  Finally, the presence of cavities and/or surface brightness fluctuations may evolve with redshift in an unknown
way, so that also the increase of the physical scale encompassed by 3 arcsec may also potentially introduce a bias.
As a result, any physical modeling is extremely complex, and on the basis of current knowledge, cannot reach a robust
description of the surface brightness distribution at the BCG position.  

Therefore, we choose to test our "flat surface brightness" assumption following a two-step procedure based on a 
phenomenological approach.  In the first step, we obtain a first assessment of the intrinsic uncertainty due to the fluctuations in 
the ICM emission based on the actual data, without modeling.  If we assume that the unresolved X-ray emission is negligible in 
all the cases where we do not detect it (in other words, if we neglect any possible sub-threshold AGN emission from the BCG), we 
can compare the noise estimate in the annulus with the "noise" in the source region.  The simplest indicator is just the ratio of the 
variance in the source region to the variance expected from the background estimate $R\equiv CTS_S/(0.213\times B)$.  This 
quantity is expected to be distributed around $R=1$ with a relative average {\sl rms} estimated as $\sqrt{(1/CTS_S+1/
(0.213\times B))}$ if our assumption of a flat surface brightness within 3 arcsec is accurate.  Under this 
assumption, we ascribe any excess variance we observe in the data to the effect of intrinsic, non-Poissonian 
fluctuations in the surface brightness at the BCG position.  Therefore, we simply multiply the statistical error by the ratio of 
the observed {\sl rms} of $R$ and the expected {\sl rms} value.  This must be regarded as a conservative, 
model-independent estimate of the uncertainty associated with the complex structure of the ICM in the inner 1.2 arcsec 
where we perform our photometry.

In the second step, we fit all our sources with a single-beta profile and a double-beta profile, after excluding the inner circle 
with a radius of 3 arcsec.  The background is then just the extrapolated surface brightness profile in the inner circle.  
The use of the information from the modeling of the entire profile except for the inner 3 arcsec will provide a different and 
independent estimate, and with respect to a fixed-aperture photometry, is not directly affected by the redshift of the source.   
When fitting a double-bet aprofile, we impose a minimum scale radius of 3 arcsec, and a maximum slope $\beta =2$, to avoid 
spurious components with extremely steep profiles.  Finally, we compare the "flat surface-brightness" values with thaoes obtained 
from single- and double-beta profile fits to investigate the presence of possible systematics that might affect our procedure.

\subsection{Soft and Hard-band Flux and Luminosity}

\noindent
For each X-ray detected BCG, we transform the observed net count rate to energy flux using the appropriate conversion factor 
at the source position, which is usually within a few arcseconds of the aimpoint of the {\sl Chandra} observation.  Conversion factors are computed 
for an average power law with slope $\Gamma = 1.8$.  Soft and hard fluxes are corrected for the Galactic absorption at the source 
position, estimated from the Galactic hydrogen map of \citet{2005LAB}.  Moreover, soft and hard fluxes measured from our 
aperture photometry are increased by 5\% and 10\% respectively, to account for the flux lost outside the aperture. Conversion 
factors in the X-ray band are computed directly to transform 2-7 keV count rates into 2-10 keV energy flux for a direct 
comparison with the literature.

We note that with aperture photometry, we compute the transmitted flux, corrected only for Galactic absorption, but not 
the intrinsic emission, which can be recovered only after accounting for the intrinsic absorption with spectral analysis. 
Since, because of the low $S/N$ and the strong ICM emission, the intrinsic absorption has, also in the best cases has a large uncertainty, 
we focus mostly on the hard-band fluxes and luminosities, where the effects of intrinsic absorption are milder.  
However, we also report  the soft-band flux, since the soft-band emission is used to establish the presence of unresolved X-ray 
emission, also in cases of non-detection in the hard band.   We will also provide a detailed spectral analysis for detected sources
in Section 4.4.

Finally, we transform the measured hard fluxes into 
rest-frame 2-10 keV luminosity by consistently applying a K-correction for a power law with slope $\Gamma = 1.8$:

\begin{equation}
L_{2-10 keV} = 4\pi D_L(z)^2 \times CF_{hard}(\Gamma=1.8, NH_{Gal}) \times S_{hard} \times K_{corr}(z)/T\, ,
\label{eq:lum}
\end{equation}

\noindent
where $D_L(z)$ is the luminosity distance computed for the seven-year WMAP cosmology, \citep{2011Komatsu}, CF$_{hard}$ 
is the conversion factor from the 2 to 7 keV to the unabsorbed 2-10 keV band, which depends on the assumed intrinsic power-law slope $\Gamma$ and the Galactic absorption, S$_{hard}$ is the hard-band photometry, $T$ is the total exposure 
time, and $K_{corr} = (1+z)^{\Gamma-2} = (1+z)^{-0.2}$ is the K correction.  
We need to compute the conversion factors at the position of each BCG regardless of its nuclear emission, 
since we require the luminosity upper limits in the hard band at the each BCG position to evaluate the depth of our 
search.  The upper limits are computed directly from the S/N threshold adopted for source detection.  These limits change 
considerably from cluster to cluster because of the different ICM emission and the different exposure time.

\section{Results\label{results}}

\subsection{Photometry and Energy Flux}
\noindent

We perform direct aperture photometry at the BCG position in the soft and hard X-ray images.  Statistical error bars are 
estimated simply as the Poisson uncertainty associated with the photon counts in the source and background regions.  
For simplicity, we will refer to all the extended emission (including the cosmic background, the instrumental noise, and the
dominant foreground ICM emission) as the "background" of our sources.  We use $bck_{flat}$ for the value obtained from 
the "flat surface brightness" assumption, and $bck_{1bfit}$ and $bck_{2bfit}$ for the values obtained from a full
surface brightness fit.  As described in Section 3.2 and Section 3.3, the measurement of $bck_{flat}$ is based 
on the simple assumption of a flat surface brightness as estimated in a ring of $1.5"<r<3"$ centered on the BCG position.  
To assess the reliability of the value $bck_{flat}$, as a first step we focus only on those sources that do not 
show unresolved emission in either of the two bands.  We also select only those that have at least seven counts in the central 
region, to have a reasonable estimate of the noise.  Then, we directly compare the value of $bck_{flat}$ with the value found
in the inner 1.2".  If the two values were statistically equivalent, we should find their ratio $R\equiv CTS_S/(0.213\times B)$ 
centered around unity with a {\sl rms}  dispersion $\sqrt{(1/CTS_S+1/(0.213\times B))}$ comparable with the typical statistical 
error.  We find that the ratio is consistent with an unity, and therefore no significant bias is found.
However, we also find that the {\sl rms}  dispersion is 16\% and 13\% higher than the statistical 
noise in the soft and hard band, respectively.  The slightly larger factor found in the soft band is expected since the most 
significant contribution to surface brightness fluctuations in the soft band is likely due to cavities in the cluster core, where 
the coldest ICM is found. On the other hand, in the hard band, the contribution of the hotter gas (typically at larger radii 
and thus less affected by cavities) is dominant.  We stress that this is a conservative upper limit to the expected noise due 
to fluctuations 
in the surface brightness of the ICM, since we are not always able to exclude sub-threshold nuclear emission, which may 
significantly contribute to the excess variance.  Therefore, we conclude that by multiplying the statistical error on $bck_{flat}$ 
by 1.16 and 1.13 in the soft and hard bands, respectively, we obtain an unbiased and robust estimate of the total uncertainty on 
the background at the position of the BCG.   

\begin{figure}
\begin{center}
\includegraphics[width=15cm]{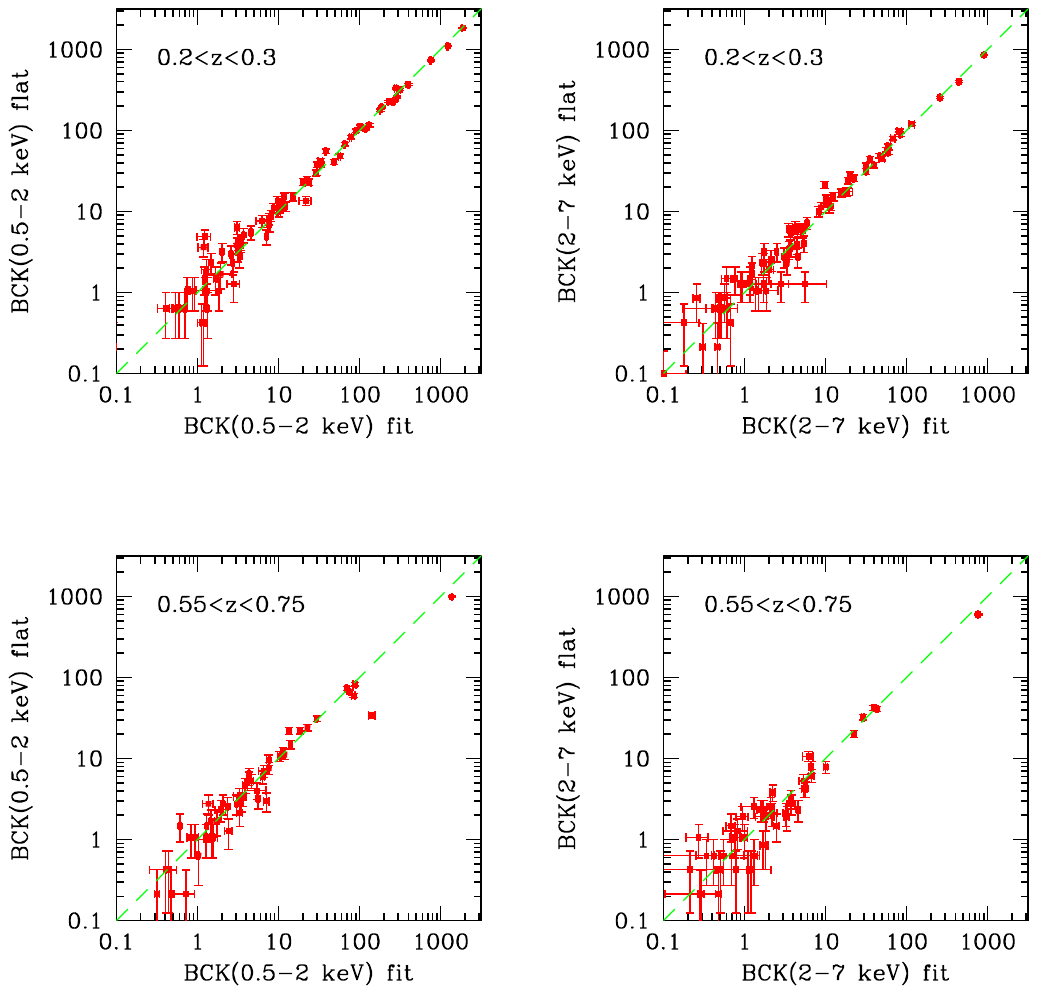}  
\caption{\label{bckfit}Comparison of the background value (total counts in the 0.5-7 keV band) assuming a flat surface brightness 
in the inner 3 arcsec ($bck_{flat}$) to the value obtained with a double-beta model fit ($bck_{2bfit}$).  The comparison is made for 
each energy band and each redshift range separately.}
\end{center}
\end{figure}

In the second step, we further investigate the robustness of our background estimate by fitting the entire surface brightness 
profile with a single-beta model profile.  All the profiles are inspected by eye and fitted with {\tt sherpa} following  the {\tt ciao} 
thread\footnote{See {\tt  http://cxc.harvard.edu/ciao/threads/radial\_profile/.}}.  We find that the values $bck_{1bfit}$ obtained with a single-beta model are on average 30\% lower than $bck_{flat}$, which may simply indicate that a single beta model is not sufficient to 
catch the rapid increase of the surface brightness in the center of a cool-core cluster.  Therefore, we repeat the
fit with a double-beta model.  The results are shown in Figure \ref{bckfit} for the soft- and hard- band images, in the low and high
redshift bins.  We find that on average, there is a good agreement within a few percent.
\footnote{We find only one source with strongly discrepant $bck_{2bfit}$ and $bck_{flat}$ values in the soft-band, high redshift 
bin.  In this case we assume the largest value of the background, obtained with the fit.  This holds in both bands and in both 
redshift intervals, showing that there are no effects related to the different physical scales sampled to estimate our background.}
By performing a direct fit of the $bck_{flat}-bck_{2bfit}$ relation, we find that in the low-redshift bin $bck_{2bfit}$ is in 
average 12\% and 8\% lower than $bck_{flat}$ in the soft and hard band, respectively, while the slope is consistent with unity 
within the errors. In the high-redshift bin, we find that $bck_{2bfit}$ is on average 10\% lower and 
11\% higher than $bck_{flat}$,  in the soft and hard band, respectively, while the slope is still consistent with unity. 
We apply this average correction to the background, and verify that the photometry is only marginally affected.  However, both methods provide values in good agreement, and at the same time, 
do not guarantee a control on the actual surface brightness fluctuations in the inner 1.2 arcsec, which still remain an
unavoidable uncertainty in this kind of study. 

\begin{figure}
\begin{center}
\includegraphics[width=8cm]{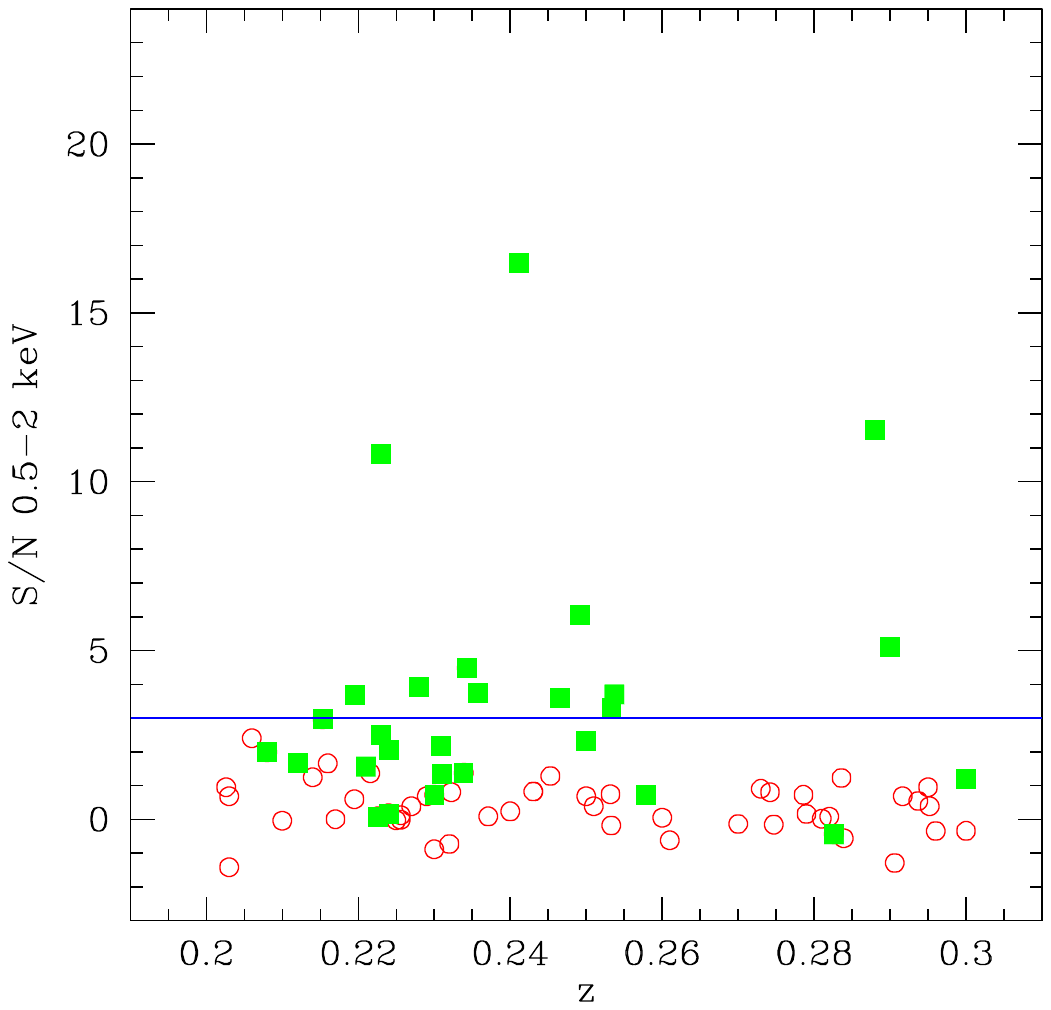}  
\includegraphics[width=8cm]{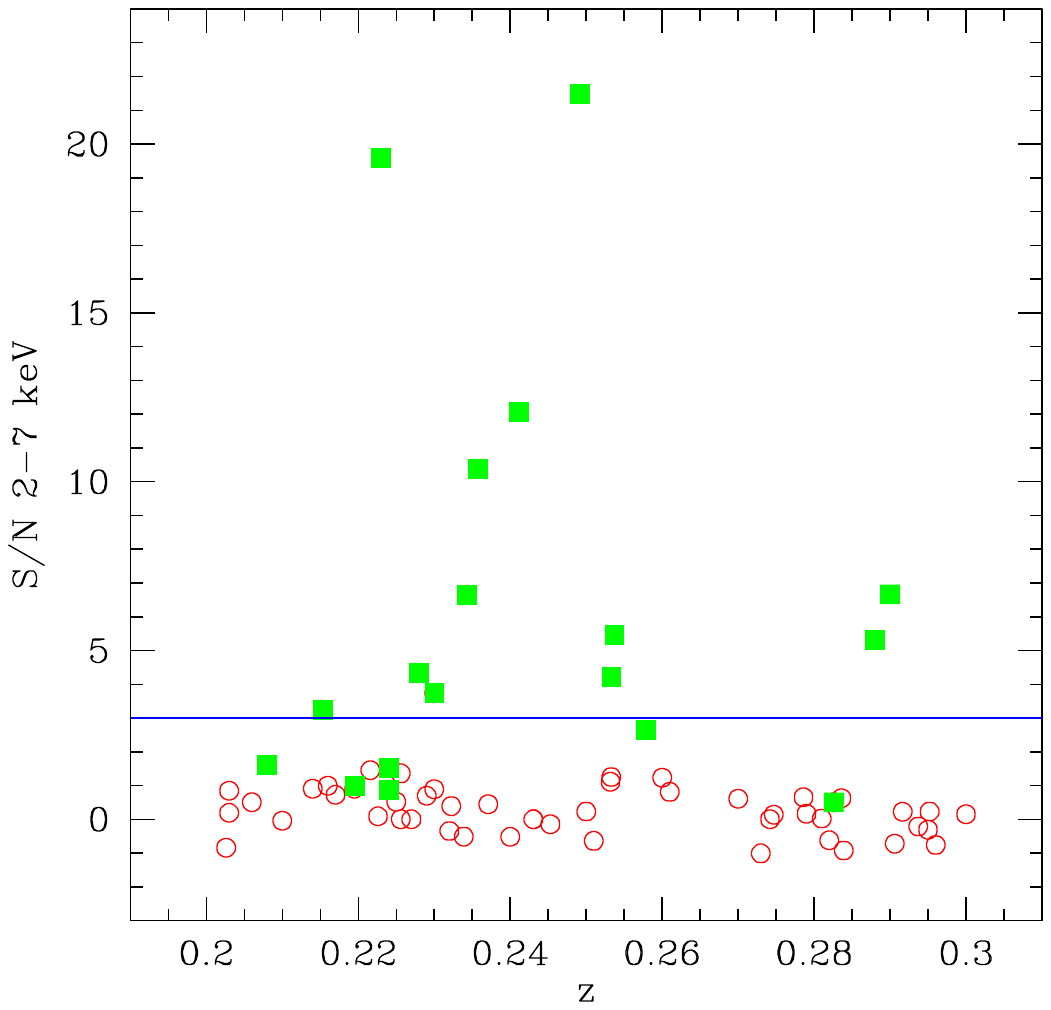} 
\caption{\label{sn_lowz}Left panel: S/N in the soft band versus redshift for the low-redshift BCG sample.  Green 
squares show BCGs with unresolved X-ray emission after visual inspection, while empty red circles are 
nondetections.  The 
horizontal line corresponds to the assumed detection threshold $S/N = 3$.   Right panel: S/N in the hard band versus 
redshift for the low-redshift BCG sample.  Symbols are the same as in the left panel.}
\end{center}
\end{figure}

\begin{figure}
\begin{center}
\includegraphics[width=8cm]{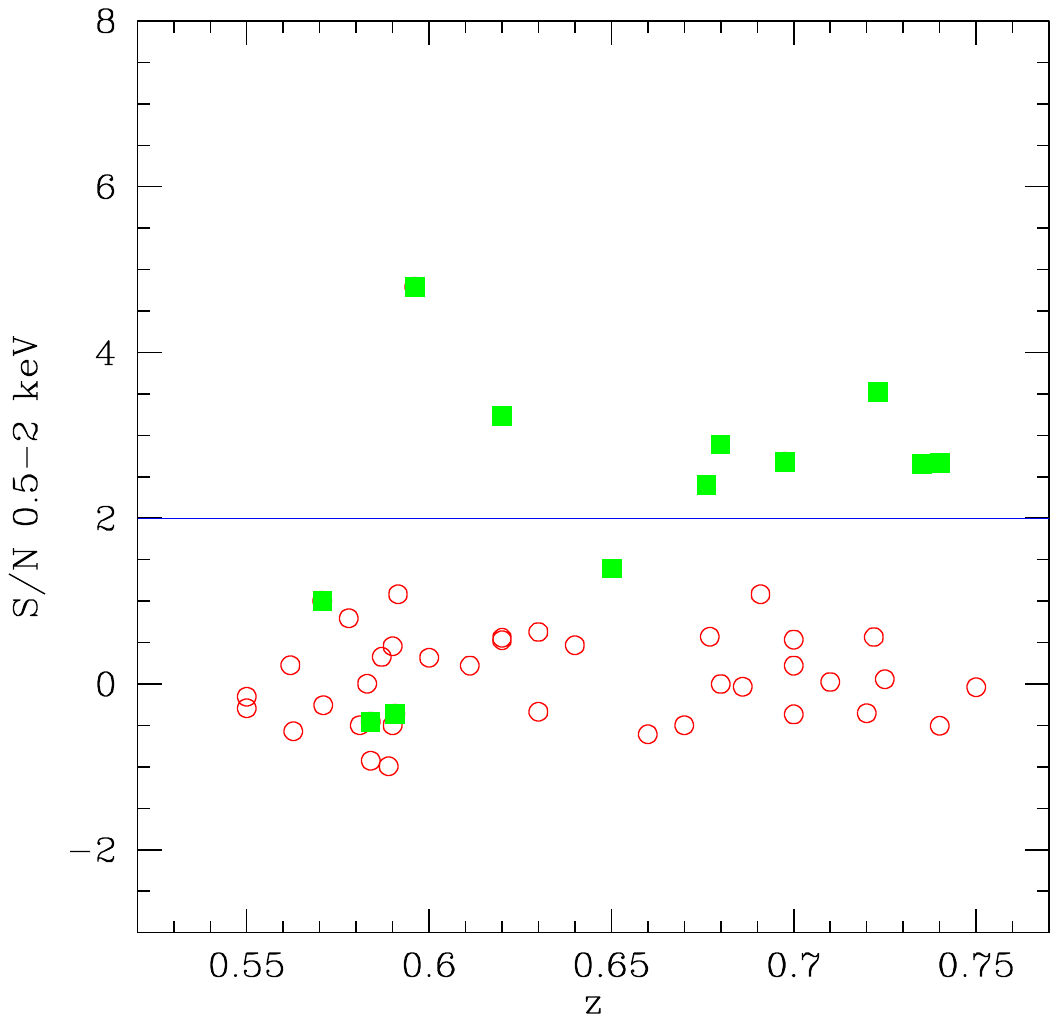}  
\includegraphics[width=8cm]{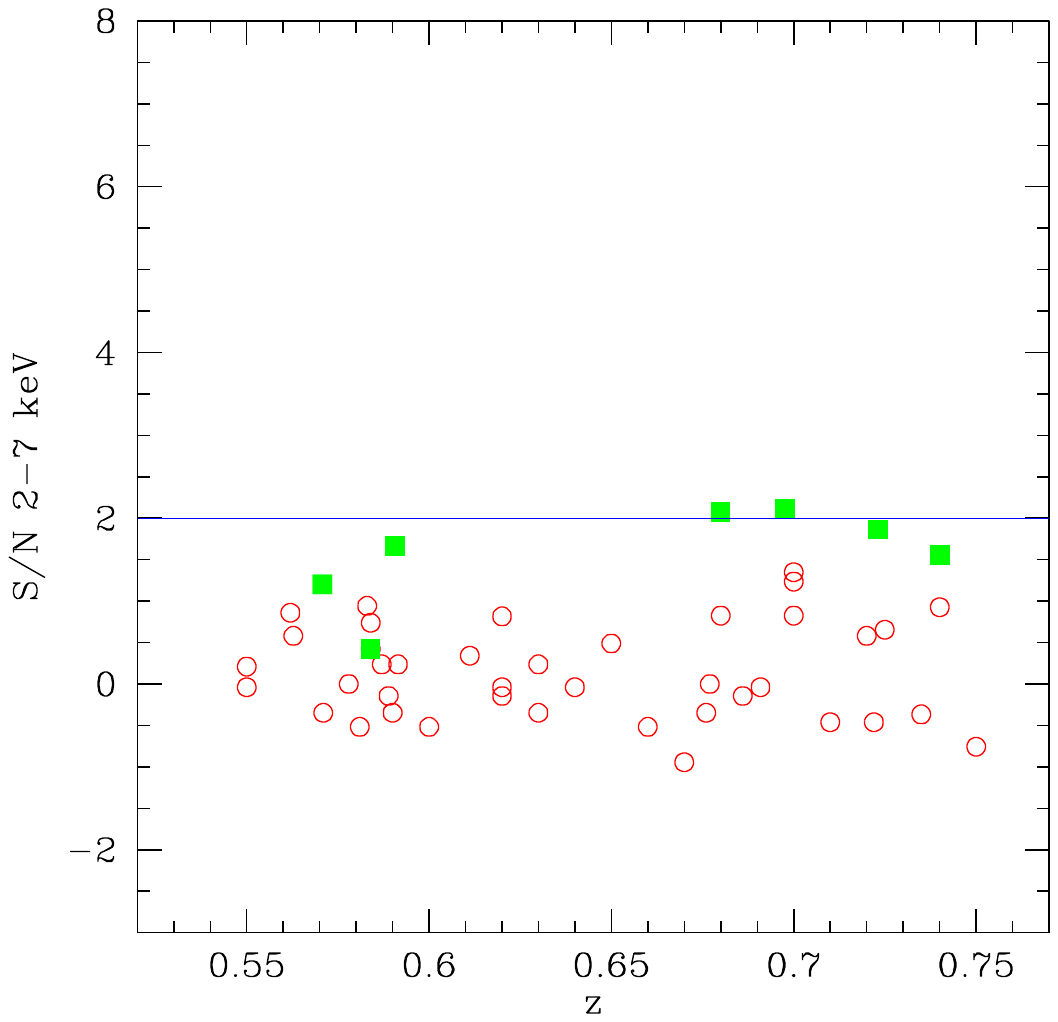} 
\caption{\label{sn_highz} Left panel: S/N in the soft band versus redshift for the high-redshift BCG sample.  Green 
solid circles show BCG with unresolved X-ray emission, while red open denote nondetections.  The horizontal line 
corresponds 
to the assumed detection threshold $S/N = 2$.   Note that the brightest source (3C254), with an $S/N\sim 56$, is not 
shown.  Right panel: S/N in the hard band versus redshift for the high-redshift BCG sample.  
Symbols as in the left panel.  Note that the two brightest sources (3C254, at $z=0.74$, and the Phoenix cluster, 
at $z=0.596$) with an $S/N \sim 38$ and $\sim 59$, respectively, are not shown.}
\end{center}
\end{figure}

In the two panels of Figure \ref{sn_lowz} we show the soft- and hard- band S/N for the low-redshift sample plotted against
the redshift.  Sources with unresolved emission detected by visual inspection at least in one band are shown with 
green squares, while sources with no apparent unresolved emission  in both bands are shown with red circles.  We note 
that the soft-band S/N distribution does not identify a clear threshold to separate sources with and without unresolved 
emission. When focusing on the low-redshift range, we find that sources with and without unresolved emission cannot be 
separated on the basis of the S/N for $S/N < 3$, while for $S/N > 3$ all the sources have been flagged with unresolved 
emission in our visual inspection.  The significant contamination at low $S/N$ is likely due to the presence of complex
structures in the cold gas, X-ray coronae, or both.  Therefore, we adopt the criterion $S/N>3$ 
in at least one band to identify sources with reliable unresolved emission among those flagged by visual 
inspection.  This threshold is shown in the panels of  Figure \ref{sn_lowz}  as a horizontal line.  This criterion identifies 14 
BCGs with unresolved nuclear X-ray emission out of 81 ($\sim$\,17\%).

For the high-redshift sample, shown in the two panels of Figure \ref{sn_highz}, the sources with  
unresolved emission are found at $S/N>2$ in the soft and hard band.  
Therefore, in this case we adopt a threshold $S/N =2$, lower than in the low-redshift sample.  This choice allows us to 
select 9 sources with visual detection and $S/N>2$ in at least one band. Therefore we have 9 BCGs with unresolved
X-ray emission out of 51, corresponding to $\sim$\,18\% of the sample, similarly to the low-redshift bin.

\begin{figure}
\begin{center}
\includegraphics[width=8cm]{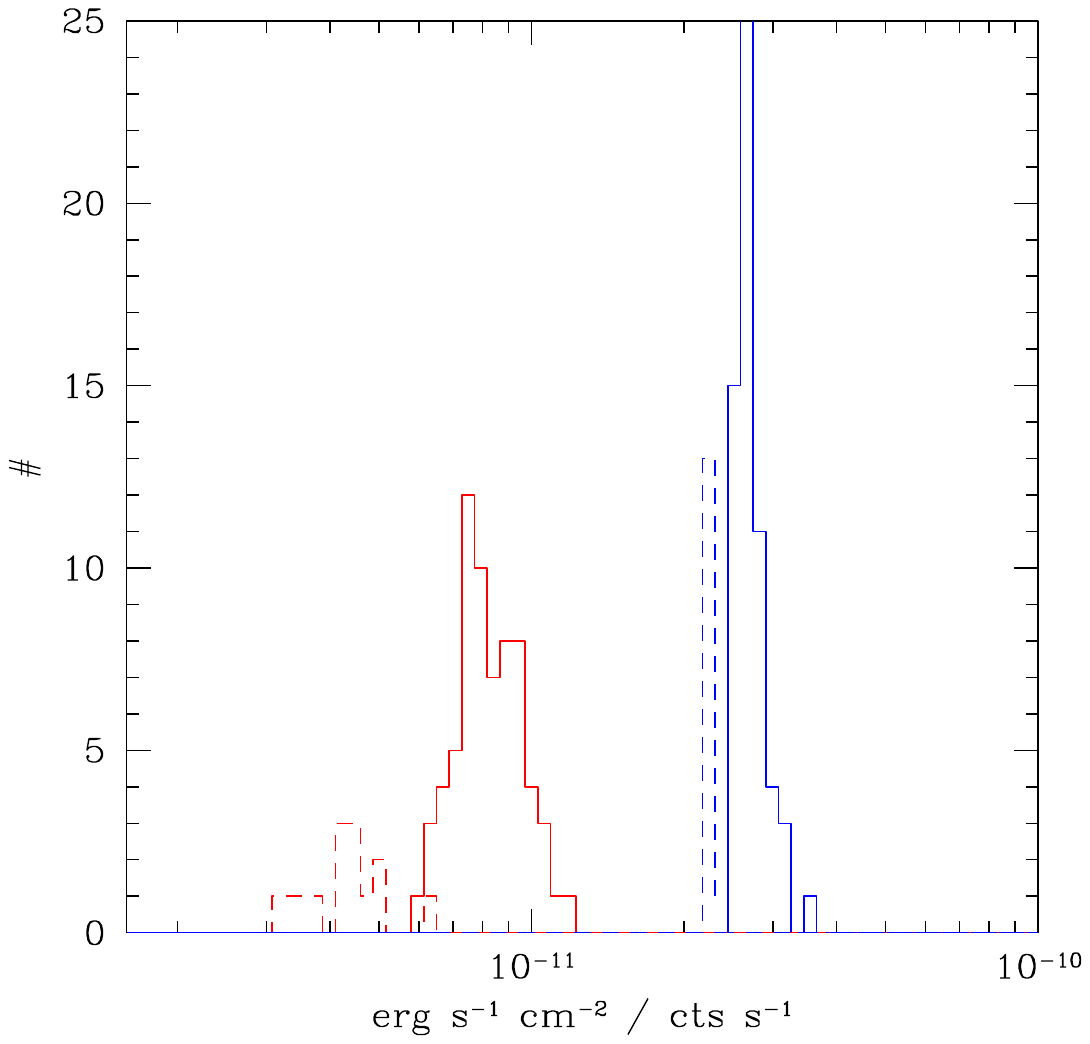}  
\includegraphics[width=8cm]{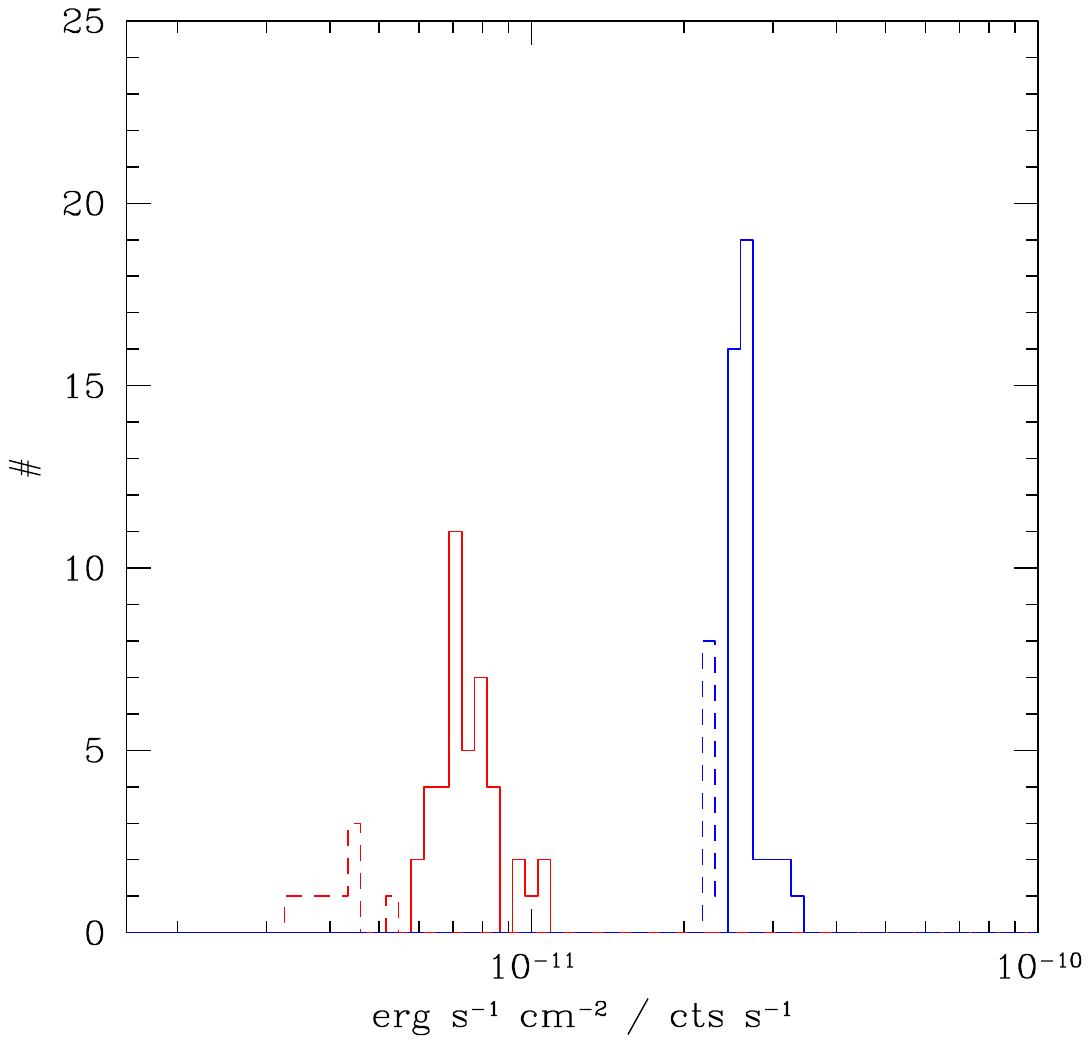}  
\caption{\label{cf_lowz}Left panel: the distributions of conversion factors in the soft band and in the hard band (from 
2-7 keV to 2-10 keV) are shown with red and blue lines for the low-z sample.  The solid lines are for ACIS-I (67 observations), 
while the dashed lines are for ACIS-S (14 observations). Right panel: same as the left panel for the high-z sample (with
40 observations with ACIS-I and 9 with ACIS-S).}
\end{center}
\end{figure}

In Figure \ref{cf_lowz} we show the distributions of the 2-7 keV count rate to 2-10 keV energy flux 
conversion factors for the soft and hard band, used to derive the
observed fluxes.  The distribution in the soft band is significantly higher than in the hard band, which is due to the
effect of the different Galactic absorption columns at the position of the BCG.  In addition, 
another source of variation is due to the mix of exposures taken at different epochs, combined with the progressive 
degradation of the effective area due to the molecular contamination of the ACIS filters over the years.  

\begin{figure}
\begin{center}
\includegraphics[width=16cm]{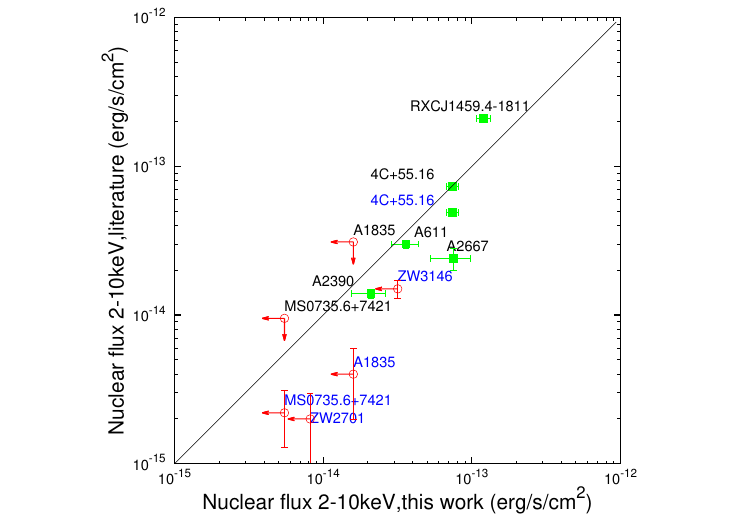}
\caption{\label{russell}Comparison between hard-band fluxes measured in this work with those measured in 
\citet{2013Russell} from spectral analysis 
and \citet{2015Hlavacek-Larrondo} from photometry, shown with black and blue labels, respectively, for the few sources
in common.  As usual, sources with unresolved emission in the soft or hard band are shown with green filled squares, 
while sources with upper limits are shown with red empty circles.  Upper limits correspond to $3\sigma$ c.l., and all 
the sources in this plot belong to the low-redshift bin.}
\end{center}
\end{figure}

In Table \ref{photom_lowz} and Table  \ref{photom_highz} we show the photometry of the BCG with unresolved X-ray 
emission in one or both bands in the  $0.2<z<0.3$ and  $0.55<z<0.75$ redshift range, respectively.  
Error bars on photometry include only the (Poissonian) statistical uncertainties, while error bars on energy fluxes 
also include the uncertainties associated with the ICM surface brightness fluctuations, as previously discussed. 
Only energy fluxes are increased by 5\% and 10\% in the soft and hard band, respectively, 
to account for the flux lost outside the extraction region.

As a check, we compare our photometric hard-band fluxes obtained with conversion factors to the values found in the literature.  
The comparison for the 7 sources in common with \citet[][from spectroscopic analysis]{2013Russell} and 5 
sources in common with \citet[][from photometry]{2015Hlavacek-Larrondo} shows a reasonable agreement, 
considering the different data reduction and the different measurement procedure (see Figure \ref{russell}).
Two sources show statistically significant differences, namely RXC J1459.4-1811 and A2667, for which 
we measure a hard-band flux about two times lower and about three times higher, respectively, than \citet{2013Russell}.  
We will comment on these two discrepant sources after we present the spectral analysis presented in Section 4.4.

We remark that in Table \ref{photom_lowz} and Table  \ref{photom_highz} we report both soft- and hard-band photometry, regardless of the detection band, so that in some cases one of the two fluxes is below the formal threshold
($S/N>3$ and $S/N>2$ in the low- and high-z sample, respectively).  In the next section, we focus only on the 
sources with a firm detection in the hard band, therefore above the selection threshold. 
This reduces the number of sources to 12 and 5 in the low- and high-z sample, respectively.

\begin{landscape}
\begin{deluxetable}{cccccccccc} 
\tablewidth{0pt}
\tablecaption{Soft- and Hard- Band Photometry of BCGs  in the  $0.2<z<0.3$ Redshift Range with Unresolved Emission Visually 
Detected and $S/N >3$ in at Least One Band. 
Conversion factors are given in units of $10^{-12}$ and $10^{-11}$ 
erg s$^{-1}$ cm$^{-2}$ / cts s$^{-1}$ in the soft and hard band, respectively. Soft and hard energy flux values
are given in units of $10^{-15}$ erg s$^{-1}$ cm$^{-2}$.  Error bars on counts include only the statistical
uncertainties, while errors on the energy flux and hard-band luminosity include also the uncertainties associated to the ICM 
surface brightness fluctuations. Hard-band luminosity is computed after equation \ref{eq:lum}.  
Fluxes and luminosities are corrected for Galactic absorption but not for intrinsic absorption. }
\tablehead{  \colhead{Cluster}  &  \colhead{Soft Net Counts}  &  \colhead{Hard Net Counts}  &  \colhead{Soft S/N}  &  \colhead{Hard S/N }  & Soft CF & Hard CF  & Soft Flux & Hard Flux & $ log(LX_{hard})$}
\startdata
\hline
RXC J1504-0248 & $ 212.75 \pm  71.61$ & $ 156.27 \pm 48.10$ &  2.97 &  3.24 &  9.186 &  2.601 & $  13.9 \pm 5.4 $ & $  30.18 \pm  10.50 $ & $ 42.59 \pm 0.13 $  \\
G256.55-65.69 & $ 33.20 \pm  8.90$ & $ 4.81 \pm 4.68$ &  3.69 &  1.01 &  8.957 &  2.688 & $  10.9 \pm 3.4 $ & $  4.96 \pm  5.46 $ & $ 41.82 \pm 0.32 $  \\
PKS 1353-341 & $ 199.25 \pm  18.30$ & $ 534.03 \pm 27.17$ &  10.83 &  19.59 &  1.083 &  2.860 & $  74.9 \pm 8.0 $ & $  555.29 \pm  31.92 $ & $ 43.89 \pm 0.02 $  \\
A2390 & $ 130.96 \pm  33.16$ & $ 79.25 \pm 18.11$ &  3.94 &  4.35 &  4.318 &  2.227 & $  6.4 \pm 1.9 $ & $  20.90 \pm  5.39 $ & $ 42.48 \pm 0.10 $  \\
A2667 & $ 8.87 \pm  12.12$ & $ 30.19 \pm 7.98$ &  0.73 &  3.74 &  3.447 &  2.188 & $  3.3 \pm 5.3 $ & $  75.33 \pm  22.49 $ & $ 43.05 \pm 0.11 $  \\
A2146 & $ 133.67 \pm  29.72$ & $ 138.44 \pm 20.70$ &  4.48 &  6.66 &  7.724 &  2.584 & $  2.9 \pm 0.7 $ & $  10.48 \pm  1.77 $ & $ 42.21 \pm 0.07 $  \\
RXC J1459.4-1811 & $ 76.23 \pm  20.28$ & $ 192.31 \pm 18.42$ &  3.74 &  10.38 &  4.900 &  2.235 & $  9.9 \pm 3.1 $ & $  119.29 \pm  12.91 $ & $ 43.27 \pm 0.04 $  \\
4C+55.16 & $ 702.57 \pm  42.52$ & $ 274.64 \pm 22.64$ &  16.49 &  12.08 &  4.140 &  2.210 & $  34.0 \pm 2.4 $ & $  74.31 \pm  6.92 $ & $ 43.09 \pm 0.04 $  \\
A2125 & $ 25.09 \pm  6.85$ & $ -0.28 \pm 1.71$ &  3.61 &  -0.16 &  6.484 &  2.594 & $  2.0 \pm 0.6 $ & $  -0.09 \pm  0.64 $ & -  \\
CL 2089 & $ 175.79 \pm  28.98$ & $ 672.77 \pm 31.22$ &  6.04 &  21.48 &  4.408 &  2.243 & $  20.0 \pm 3.8 $ & $  408.38 \pm  21.41 $ & $ 43.86 \pm 0.02 $  \\
RXC J1023.8-2715 & $ 80.95 \pm  24.36$ & $ 62.81 \pm 14.78$ &  3.31 &  4.22 &  4.605 &  2.234 & $  10.8 \pm 3.8 $ & $  42.44 \pm  11.29 $ & $ 42.89 \pm 0.10 $  \\
CL 0348 & $ 102.68 \pm  27.63$ & $ 80.84 \pm 14.70$ &  3.70 &  5.46 &  4.401 &  2.224 & $  9.7 \pm 3.0 $ & $  40.58 \pm  8.34 $ & $ 42.87 \pm 0.08 $  \\
A611 & $ 236.73 \pm  20.40$ & $ 53.27 \pm 9.90$ &  11.55 &  5.33 &  3.809 &  2.200 & $  26.5 \pm 2.6 $ & $  36.09 \pm  7.58 $ & $ 42.95 \pm 0.08 $  \\
3C438 & $ 102.87 \pm  20.06$ & $ 106.92 \pm 15.94$ &  5.10 &  6.66 &  6.265 &  2.311 & $  4.3 \pm 1.0 $ & $  17.17 \pm  2.89 $ & $ 42.63 \pm 0.07 $  \\
\hline
\enddata
\label{photom_lowz}

\end{deluxetable}
\end{landscape}

\begin{landscape}
\begin{deluxetable}{cccccccccc} 
\tablewidth{0pt}
\tablecaption{Soft- and Hard Band Photometry of BCGs with Unresolved Emission Detected with $S/N >2$ in at Least One Band, 
in the  $0.55<z<0.75$ Redshift Range.
Conversion factors are given in units of $10^{-12}$ and $10^{-11}$ 
erg s$^{-1}$ cm$^{-2}$ / cts s$^{-1}$ in the soft and hard band, respectively. Soft and hard energy flux values
are given in units of $10^{-15}$ erg s$^{-1}$ cm$^{-2}$.  Error bars on counts include only the statistical
uncertainties, while errors on the energy flux and hard-band luminosity include also the uncertainties associated to the ICM 
surface brightness fluctuations. Hard-band luminosity is computed after equation \ref{eq:lum}. 
Fluxes and luminosities are corrected for Galactic absorption but not for intrinsic absorption.}
\tablehead{  \colhead{Cluster}  &  \colhead{Soft Net Counts}  &  \colhead{Hard Net Counts}  &  \colhead{Soft S/N}  &  \colhead{Hard S/N }  & Soft CF & Hard CF  & Soft Flux & Hard Flux & $ log(LX_{hard})$}
\startdata
\hline
SPT-CL J2344-4243 & $ 263.33 \pm  54.86$ & $ 4638.24 \pm 84.55$ &  4.79 &  54.79 &  1.006 &  2.996 & $  21.6 \pm 5.2 $ & $  1184.35 \pm  24.40 $ & $ 45.19 \pm 0.01 $  \\
RCS 1419+5326 & $ 33.60 \pm  10.27$ & $ 3.31 \pm 3.95$ &  3.24 &  0.82 &  3.886 &  2.219 & $  2.4 \pm 0.9 $ & $  1.43 \pm  1.94 $ & $ 42.31 \pm 0.37 $  \\
ACT J0206-0114 & $ 9.93 \pm  4.03$ & $ -0.21 \pm 0.51$ &  2.41 &  -0.35 &  8.207 &  2.527 & $  2.9 \pm 1.4 $ & $  -0.20 \pm  0.54 $ &  -  \\
SDSS J1004+4112 & $ 33.57 \pm  11.51$ & $ 12.68 \pm 6.01$ &  2.89 &  2.08 &  4.566 &  2.217 & $  0.7 \pm 0.3 $ & $  1.27 \pm  0.68 $ & $ 42.36 \pm 0.19 $  \\
MACS 0744.8+3927 & $ 42.39 \pm  15.72$ & $ 22.00 \pm 10.29$ &  2.68 &  2.12 &  7.160 &  2.581 & $  3.6 \pm 1.5 $ & $  7.03 \pm  3.72 $ & $ 43.13 \pm 0.18 $  \\
SPT-CL J2043-5035 & $ 55.28 \pm  15.59$ & $ 15.73 \pm 8.33$ &  3.52 &  1.87 &  7.918 &  2.611 & $  5.8 \pm 1.9 $ & $  5.72 \pm  3.42 $ & $ 43.07 \pm 0.20 $  \\
RCS 1107.3-0523 & $ 19.04 \pm  7.05$ & $ -0.71 \pm 1.83$ &  2.66 &  -0.37 &  4.259 &  2.253 & $  0.9 \pm 0.4 $ & $  -0.19 \pm  0.55 $ & - \\
3C254 & $ 3224.95 \pm  66.61$ & $ 1502.19 \pm 43.30$ &  48.34 &  34.61 &  3.369 &  2.176 & $  386.2 \pm 9.3 $ & $  1217.29 \pm  39.65 $ & $ 45.42 \pm 0.01 $  \\
SPT-CL 0001-5748 & $ 25.52 \pm  9.46$ & $ 8.53 \pm 5.38$ &  2.67 &  1.56 &  7.383 &  2.613 & $  6.6 \pm 2.8 $ & $  8.14 \pm  5.79 $ & $ 43.25 \pm 0.23 $  \\
\hline
\enddata
\label{photom_highz}

\end{deluxetable}
\end{landscape}

\subsection{The Fraction of X-Ray Luminous BCG as a Function of $L_X$\label{xlf}}

The fraction of X-ray emitting BCGs above a given X-ray luminosity is computed as

\begin{equation}
F_{BCG}(>L_X) = \Sigma_{>L_X} {1\over  {N(L_{up}<L_X)}}
\end{equation}

\noindent
where the sum is computed for all the BCGs with a hard band X-ray luminosity higher than $L_X$, and $N(L_{up}<L_X)$ is the 
number of clusters for which the luminosity corresponding to the detection threshold is lower than $L_X$, in other words, all 
the clusters where we should have seen the AGN in the BCG if above $L_X$.  Given our selection threshold $S/N>3$ and
$S/N>2$  in the hard band for the low- and high-z sample, respectively, we have a well-defined detection threshold in 
hard-band luminosity at each BCG position.  This value defines the completeness of our sample in luminosity.  Clearly, the 
completeness correction mostly affects the lowest luminosity bins, and the correction is more important at higher redshift.

\begin{figure}
\begin{center}
\includegraphics[width=14cm]{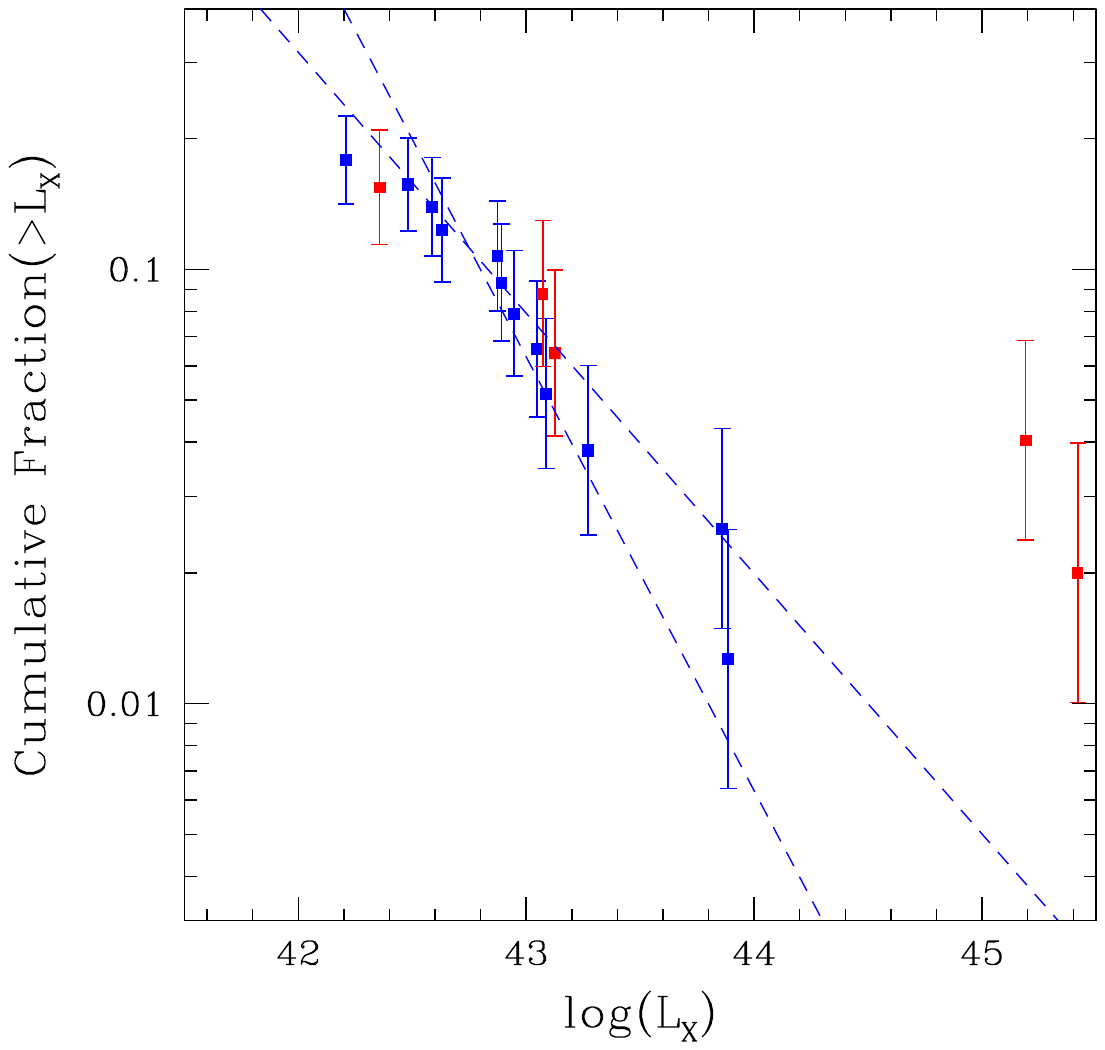}
\caption{\label{XLF}Cumulative fraction of X-ray luminous BCGs as a function of the hard-band $L_X$ in the redshift range 
$0.2<z<0.3$ (blue squares) and $0.55<z<0.75$ (red squares).  The two dashed lines bracketing the cumulative luminosity 
distribution in the low-redshift bin, have slopes of $-0.6$ and $-1$.}
\end{center}
\end{figure}

The cumulative fractions of X-ray luminous BCGs as a function of $L_X$ in the low- and high-redshift bins are shown in 
Figure \ref{XLF}. Error bars are the Poissonian error bars due to the finite numbers, so that roughly 
$\sigma = \sqrt{(N(L_{BCG}>L_X)}/N(L_{up}<L_X)$, where $N(L_{BCG}>L_X)$ is the number of BCGs with a hard-band 
luminosity higher than $L_X$. In both samples, the lowest luminosity detected is $\geq 10^{42}$ erg s$^{-1}$.  The average 
slope of the cumulative fraction in the low-z bin is between $\sim -0.6$ and $\sim -1$, with a 
very weak hint of a steeper slope at $L_X \geq 10^{43}$ erg s$^{-1}$.  The limited statistics in 
the high-redshift bin, where we have only five sources, prevents us from drawing any conclusion on the slope.  
However, we are able to establish that the normalization of the X-ray luminosity function in the  Seyfert-like luminosity 
range ($L_X<10^{44}$ erg s$^{-1}$) is consistent with the low-z sample, while a striking difference is given by the 
presence of two extremely luminous quasars (in the BCG of the Phoenix cluster and 3C254) that are completely absent at 
low redshift. Taken at face value, the measured fraction of X-ray luminous BCGs in our sample points toward no evolution 
below $10^{44}$ erg s$^{-1}$ and a possible evolution above this value.  This result is in broad agreement with 
\citet{2013Hlavacek-Larrondo}, where they find significant positive evolution with redshift.  However, their results were 
based on a sample of X-ray bright clusters with strong cavities, while we aim at including the widest range of halo masses 
and environment offered by the {\sl Chandra} archive. Clearly, any conclusion on evolution must await the use of the entire
{\sl Chandra} archive, with the same strategy as was used in this work. Eventually, on the basis of a larger statistics, we will explore the 
X-ray luminosity function of BCGs in subsamples extracted from complete and well-defined cluster catalogs.

\begin{figure}
\begin{center}
\includegraphics[width=8cm]{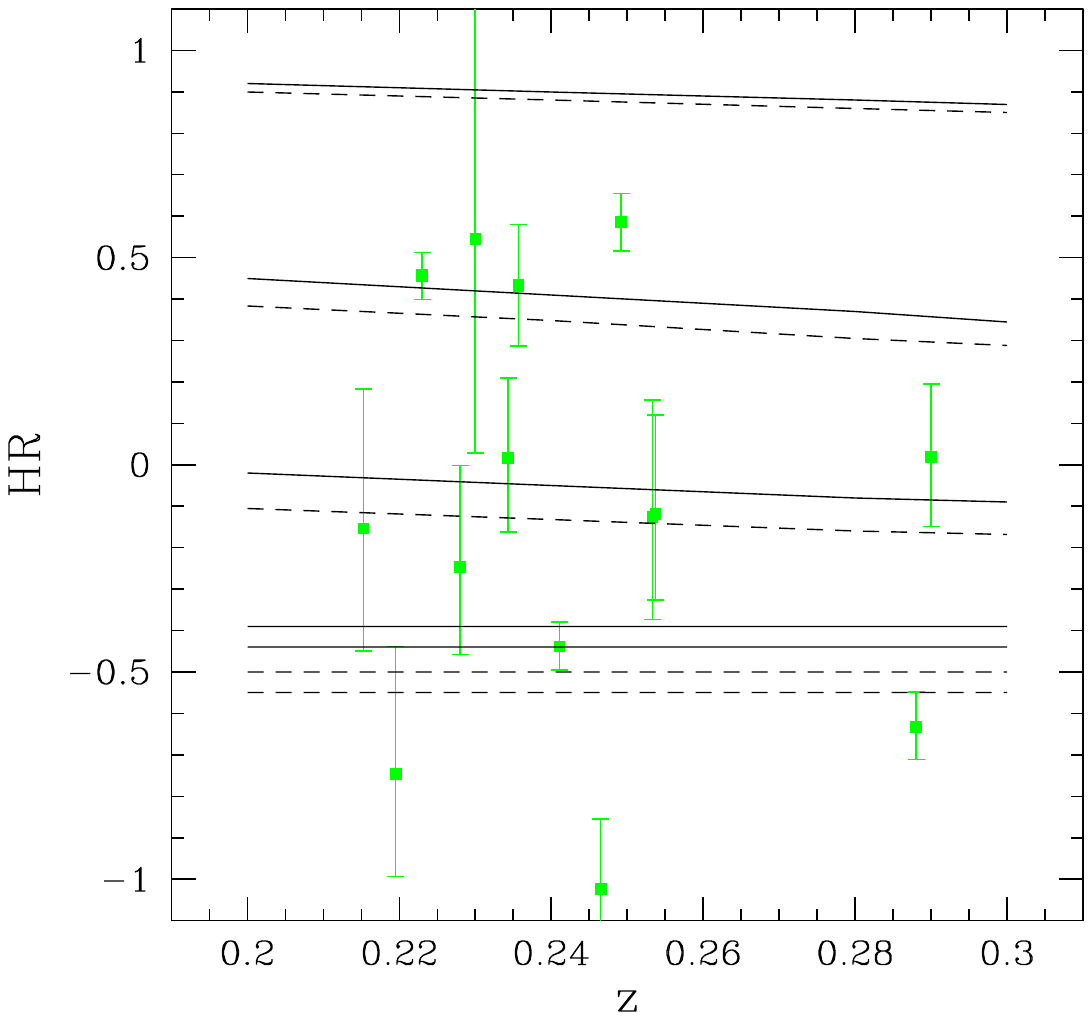}
\includegraphics[width=8cm]{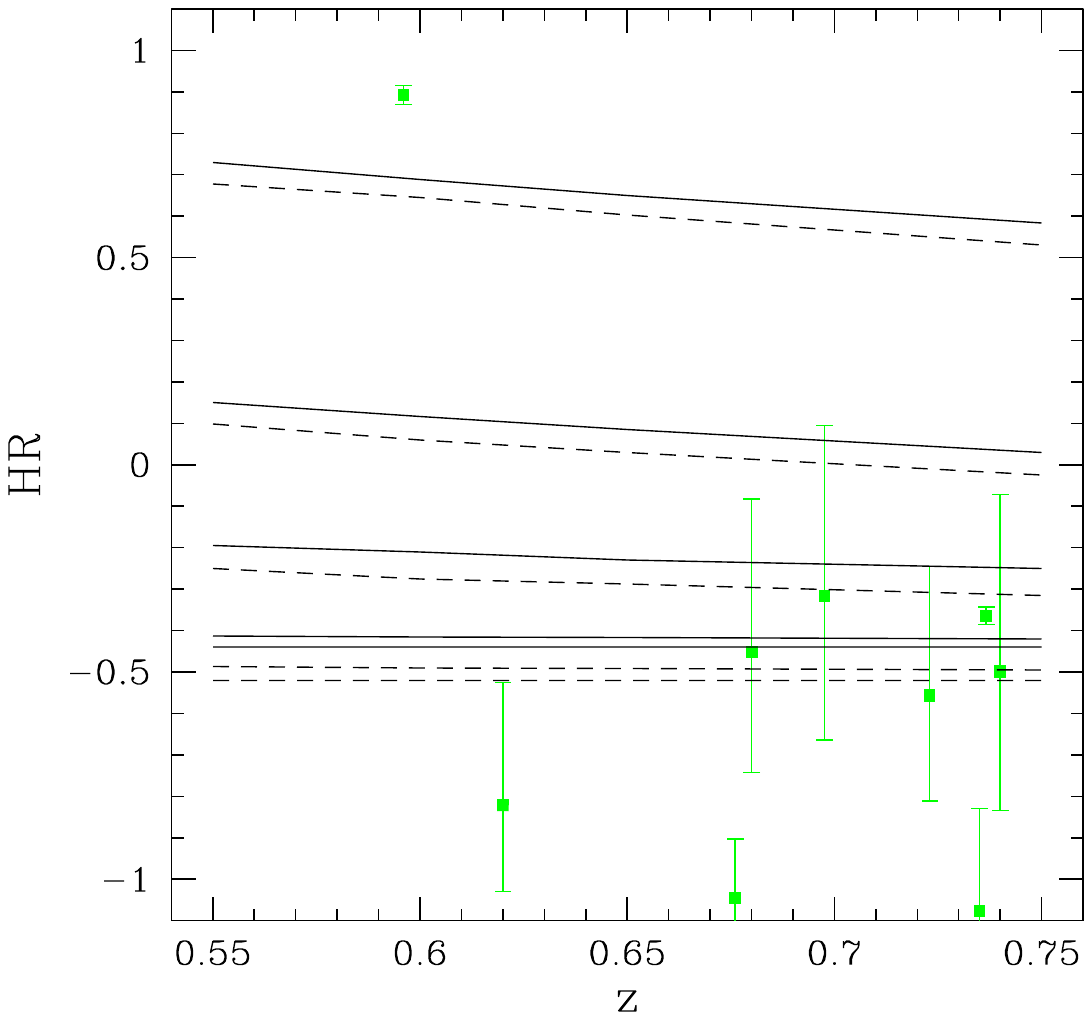}
\caption{\label{hr}Left panel: hardness ratio for the sources with unresolved X-ray emission in the soft 
or hard band, in the redshift range 
$0.2<z<0.3$.  Erorr bars correspond to 1$\sigma$ confidence level.  Solid (dashed) lines correspond to typical hardness ratios 
measured with ACIS-I (ACIS-S) for an intrinsic equivalent hydrogen-absorbing column of (from bottom to top) $10^{20}$, 
 $10^{21}$, $10^{22}$, $3\times 10^{22}$, and $10^{23}$ cm$^{-2}$.  
 Right panel: same as in the left panel for the sources in the $0.55<z<0.75$ redshift range.}
\end{center}
\end{figure}

\begin{figure}
\begin{center}
\includegraphics[width=8cm]{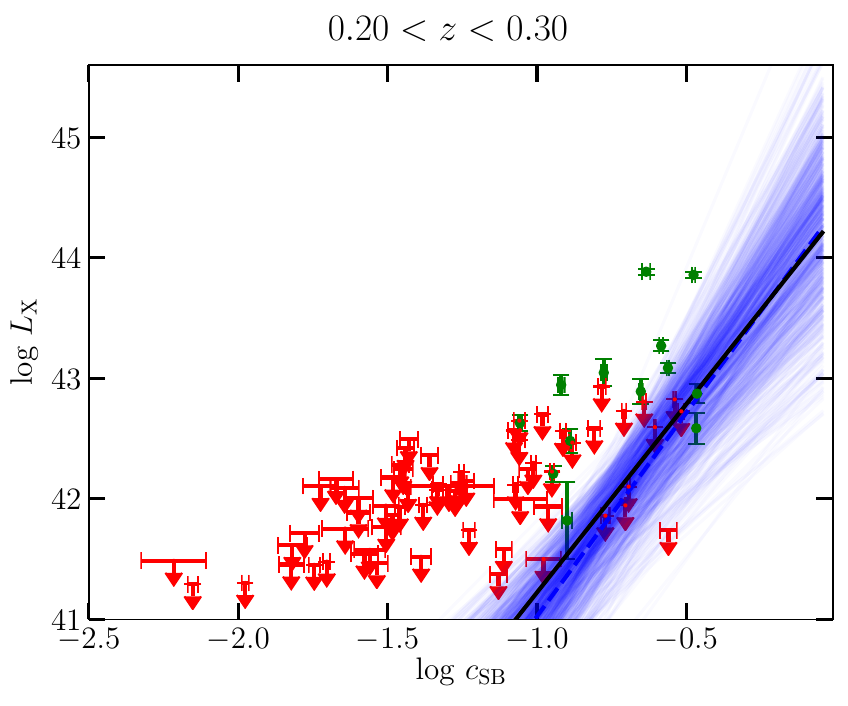}
\includegraphics[width=8cm]{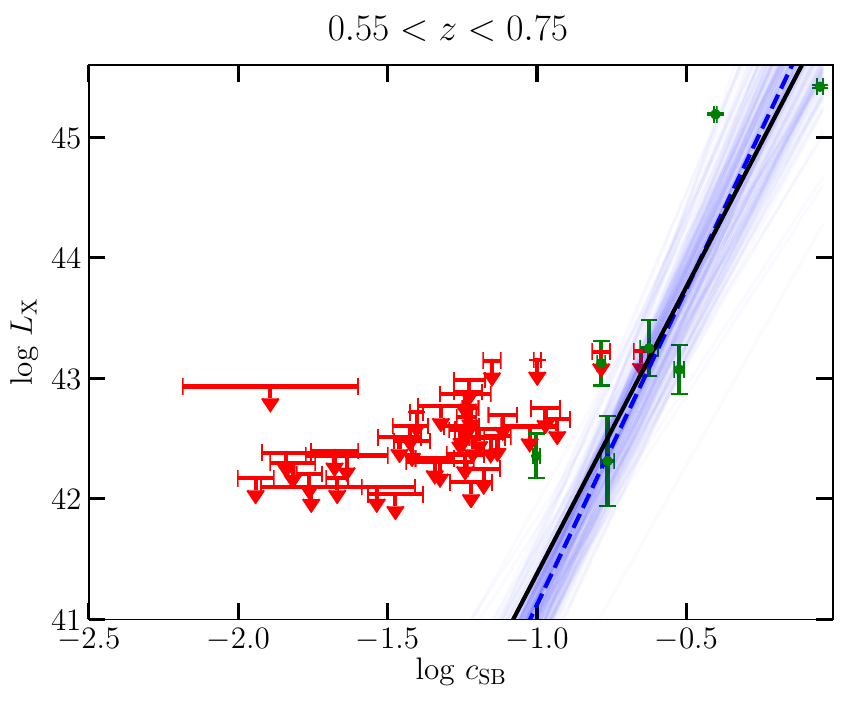}
\caption{\label{concentration}Left panel: hard-band luminosity versus the BCG concentration parameter in the 
$0.2<z<0.3$ redshfit range.  Sources with positive photometry in the hard band and $S/N>3$ in at least one band are shown 
with green solid squares.  Red arrows shows the $3\sigma$ upper limits for the BCGs with no nuclear emission. Error bars 
correspond to the 1$\sigma$ confidence level.  Black solid and blue dashed lines are the best fit obtained from a censored-data
analysis using the software ASURV and LINMIX\_ERR, respectively.  The light blue lines represent 400 different realizations of the 
$\log L_{\rm X}-\log c_{SB}$ relation from LINMIX\_ERR.  Right panel: same as in the left panel for the sources with positive 
photometry in the hard band and $S/N>2$ in the $0.55<z<0.75$ redshift range.}
\end{center}
\end{figure}

\subsection{Average Spectral Properties and Connection with Cool Cores}

For a first-cut evaluation of the spectral properties of the X-ray emitting BCGs, we compute their hardness ratio, simply defined as 
$HR\equiv (C_{hard}-C_{soft})/(C_{hard}+C_{soft})$, where $C_{hard}$ and $C_{soft}$ are the source net counts measured in the 
hard and soft band, respectively, and corrected for vignetting. In Figure \ref{hr}, left panel, we show the hardness ratios for the
sources with unresolved X-ray emission in at least one of the two bands in the low-redshift bin. We also plot solid (dashed) lines 
corresponding to the typical hardness ratio measured with ACIS-I (ACIS-S) for an intrinsic equivalent hydrogen-absorbing 
column of (from bottom to top) $10^{20}$, $10^{21}$, $10^{22}$, $3\times 10^{22}$, and $10^{23}$ cm$^{-2}$.  These 
representative curves are computed for a typical {\sl Chandra} observation at the aimpoint for a spectrum with an intrinsic 
emission described by a power law of $\Gamma=1.8$, considering an average Galactic absorbing column of 
$3\times 10^{20}$ cm$^{-2}$.  We note that roughly half of the sample in the low-redshift bin shows hints of intrinsic absorption 
($HR\geq 0$, corresponding roughly to $10^{22}$ cm$^{-2}$) in the soft band.  This implies that to compute the total intrinsic 
X-ray luminosity, we need to correct for intrinsic absorption below 2 keV.     
In Figure \ref{hr}, right panel, we show the hardness ratio for the sources with unresolved X-ray emission in at least one of the two 
bands in the high-redshift sample.  Only one source is clearly absorbed \citep[SPT-CL J2344-4243, see, e.g. ][]{2015Tozzib}, 
while the other sources are consistent with the spectrum of unabsorbed AGN ($HR\sim -0.5$). 

We also compute the concentration parameter (defined as the ratio of the energy flux 
in the soft band within 40 kpc to that within 400 kpc) at the BCG position for all our groups and clusters.  
The two fluxes are obtained after removing unresolved emission, including the central AGN when present.
Our definition of the concentration parameter is different from that of \citet{2008Santos}, which 
is computed at the peak of the X-ray surface brightness.  Clearly, the two definitions agree only when the BCG is 
located precisely at the maximum of the diffuse X-ray emission.  In Figure \ref{concentration} we show the
measured hard band luminosity for the sources with positive hard band photometry detected at least in one band
in the low and high redshift bins.   We find that, on one hand, BCGs with nuclear emission are preferentially in 
stronger cores, with concentration parameter $c_{SB}>0.1$.  On the other hand, only one-third of the clusters with 
$c_{SB}>0.1$ host an AGN with $L_X>10^{42}$ erg s$^{-1}$ in the BCG.  For example, we do not find nuclear activity in 
MS 0735.6+7421, which hosts a strong cool core and is one of the most powerful mechanical outburst  known to date 
\citep{2005McNamara,2007Gitti}, as already shown in Figure \ref{HST_images}.  One may argue that some level of nuclear 
X-ray emission may be present in all the strong cool cores, possibly hidden by the overwhelming ICM emission. 
To explore this possibility and the effects of the many upper limits, we perform a censored-data analysis on the
$log(L_X)$-$log(c_{SB})$ relation.  Owing to the large number of upper limits, we are aware that we are dealing with an extreme
situation, and the results should be critically assessed before drawing any conclusion.
We adopt the LINMIX\_ERR software \footnote{This algorithm has been implemented in Python and its description can be found at 
http://linmix.readthedocs.org/en/latest/src/linmix.html.} \citep{2007Kelly}.  This method
accounts for measurement errors on both independent and dependent variable, nondetections, and intrinsic scatter by adopting a 
Bayesian approach to compute the posterior probability distribution of parameters, given observed data.
This has been argued to be among the most robust 
regression algorithms with the possibility of reliable estimation of intrinsic random scatter on the regression. 
We also consider the Astronomy Survival Analysis software package (ASURV rev. 1.2; \citealt{isobe90}; \citealt{lavalley92}), which is 
widely used in the literature. ASURV implements the bivariate data-analysis methods and also properly treats censored data using 
the survival analysis methods (\citealt{feigelsonnelson85}; \citealt{isobe86}). We have employed the full parametric estimate and 
maximized 
regression algorithm to perform the linear regression of the data.  The results are shown in Figure \ref{concentration} with a
continuous and dashed line, from the ASURV and LINMIX\_ERR analysis, respectively.  For the low-redshift sample, we find a 
slope $\sim 3$, while in the high-redshift sample the slope is even steeper $\geq 4$.
Moreover, at low redshift, we find a low normalization, driven by the many upper limits at $c_{SB}>0.1$, while at high
redshift the normalization is driven by the detections, given the very low number of upper limits at $c_{SB}>0.1$.
The main conclusion we can reach from our analysis is that AGN with $L_X>10^{42}$ erg s$^{-1}$  ($L_X>10^{43}$ 
erg s$^{-1}$) appear only above $c_{SB}>0.1$ in the low-(high-) redshift range.  In addition, above the same X-ray luminosity
threshold, AGN do not sit in non-cool-core cluster ($c_{SB}<0.11$).   

As we have discussed, spectral analysis may be helpful in identifying nonthermal emission, possibly associated with a 
central AGN, through the measurement of spectra harder than expected from the thermal ICM, as has been proposed 
in \citet{2013Hlavacek-Larrondo}.   However, as explained in Section 3.2, this type of diagnostic based on spectral 
shape needs a very high S/N, and therefore is not suitable for exploring the low-luminosity range.  Therefore, we limit our 
spectral analysis to the sources with unresolved emission detected with our photometry, as described in the next section.

\subsection{X-Ray Spectral Analysis}

We perform a standard spectral analysis on the sources listed in Table \ref{photom_lowz} and \ref{photom_highz}
using a simple physical model consisting in an absorbed power law plus a local Galactic absorption ({\tt Xspec} model
{\tt tbabs $\times$ zwabs $\times$ pow}).  We extract source and background spectra from the same extraction regions as we used for 
photometry.   Calibration files are the same used to compute the conversion factors.
Our spectral analysis is therefore based on the same background subtraction used in our photometry.  
Our aim is to confirm our results and explore the distribution of intrinsic absorption. However, we remark that
spectral analysis in these extreme conditions of strong background can have a complex effect on the best-fit values
of the spectral parameters.  A proper approach would require the combined analysis of an absorbed power law plus
a thermal component at the same time. Clearly, this is feasible
only for very bright sources because of the strong degeneracy of a composite model. The spectral analysis discussed in this work should therefore be simply regarded as 
an extension of our photometric study.

\subsubsection{Spectral Analysis of Sources at $0.2<z<0.3$}

\begin{figure}
\begin{center}
\includegraphics[width=8cm]{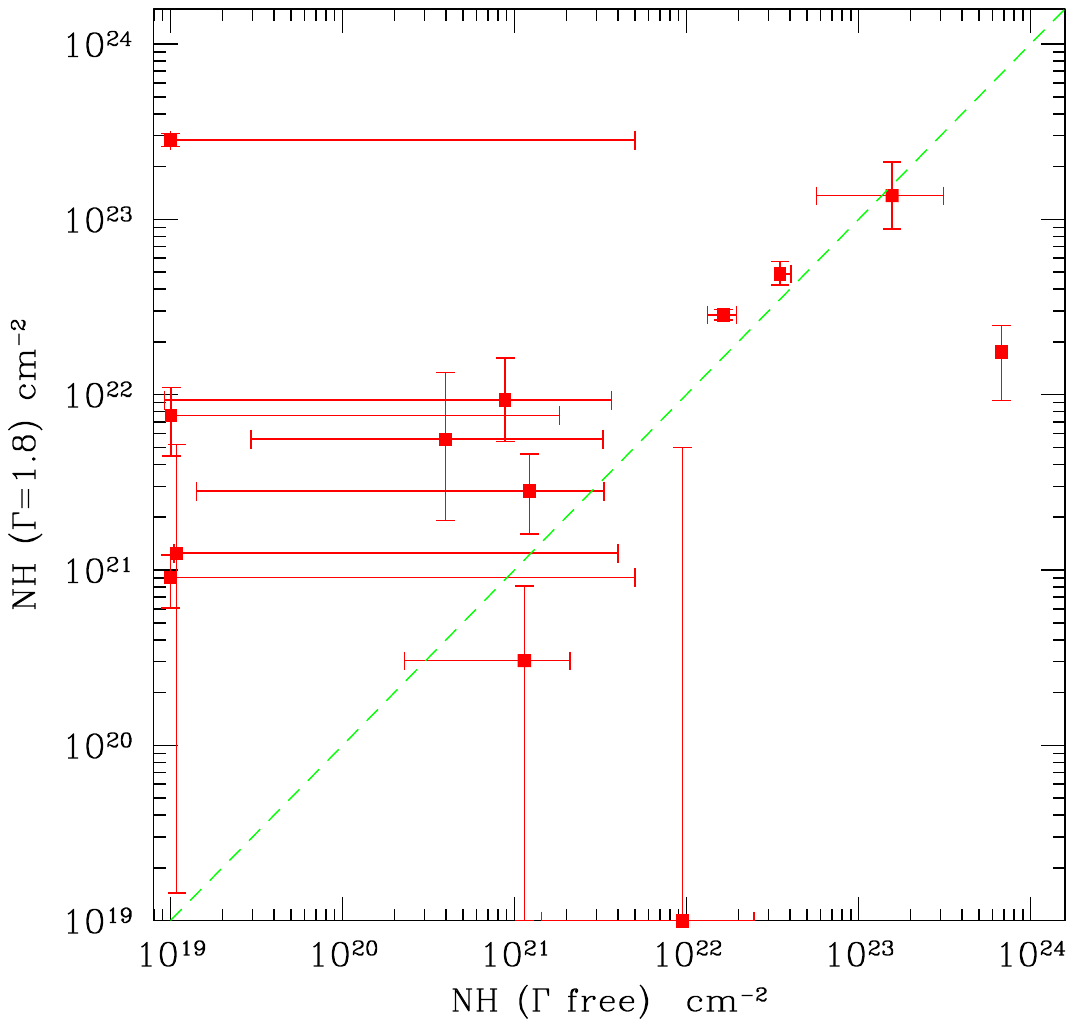} 
\includegraphics[width=8cm]{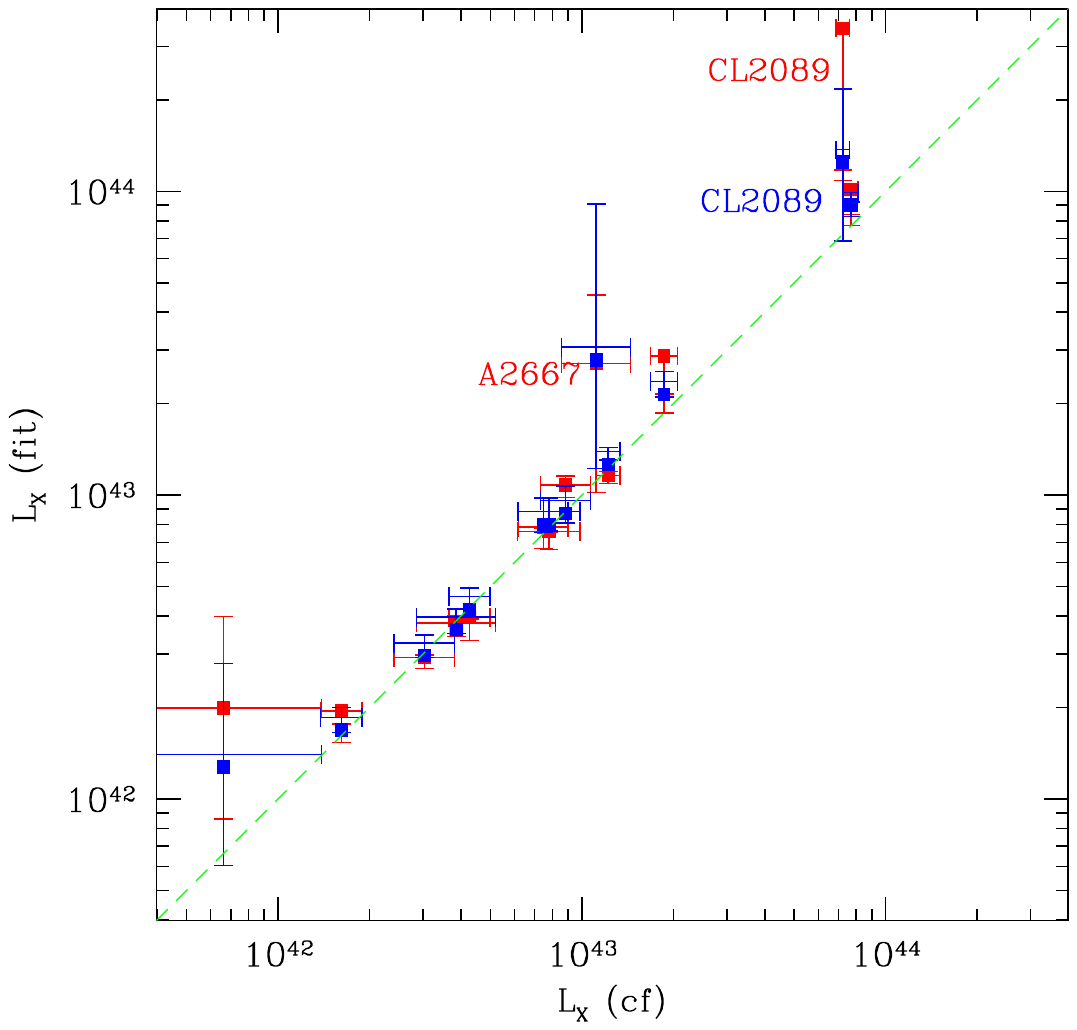} 
\caption{\label{fit_vs_cf}Left panel: comparison of best-fit values for intrinsic absorption $N_H$ obtained with a free spectral 
slope, and with a slope frozen to $\Gamma=1.8$ for the sources in the low-redshift sample.  
Right panel: rest-frame 2-10 keV, unabsorbed luminosity obtained from 
spectral analysis, compared to the value obtained from aperture photometry, and not 
corrected for intrinsic absorption.  Values obtained with a free spectral slope are shown in blue, while those obtained for 
$\Gamma=1.8$ are shown in red.  }
\end{center}
\end{figure}

\begin{figure}
\begin{center}
\includegraphics[width=8cm]{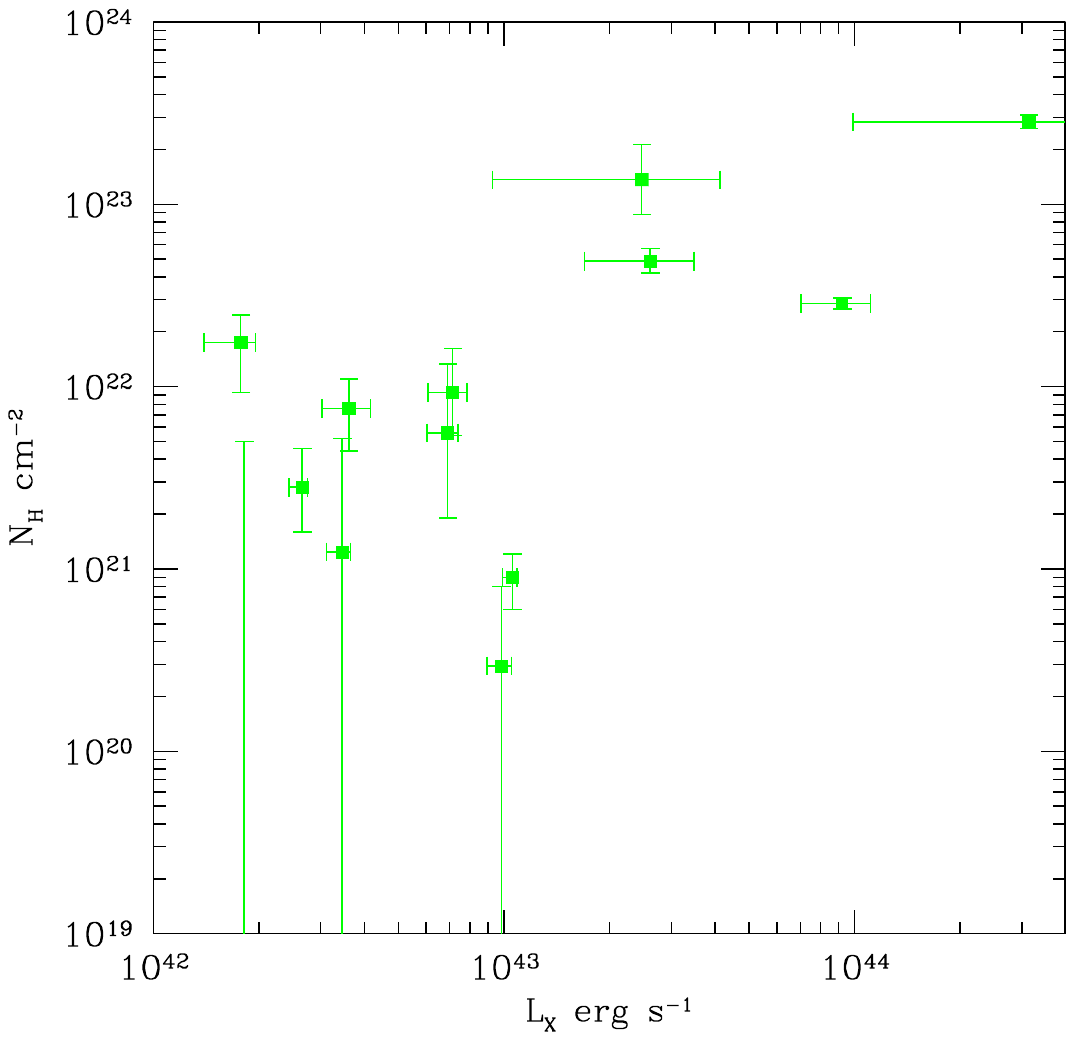}
\includegraphics[width=8cm]{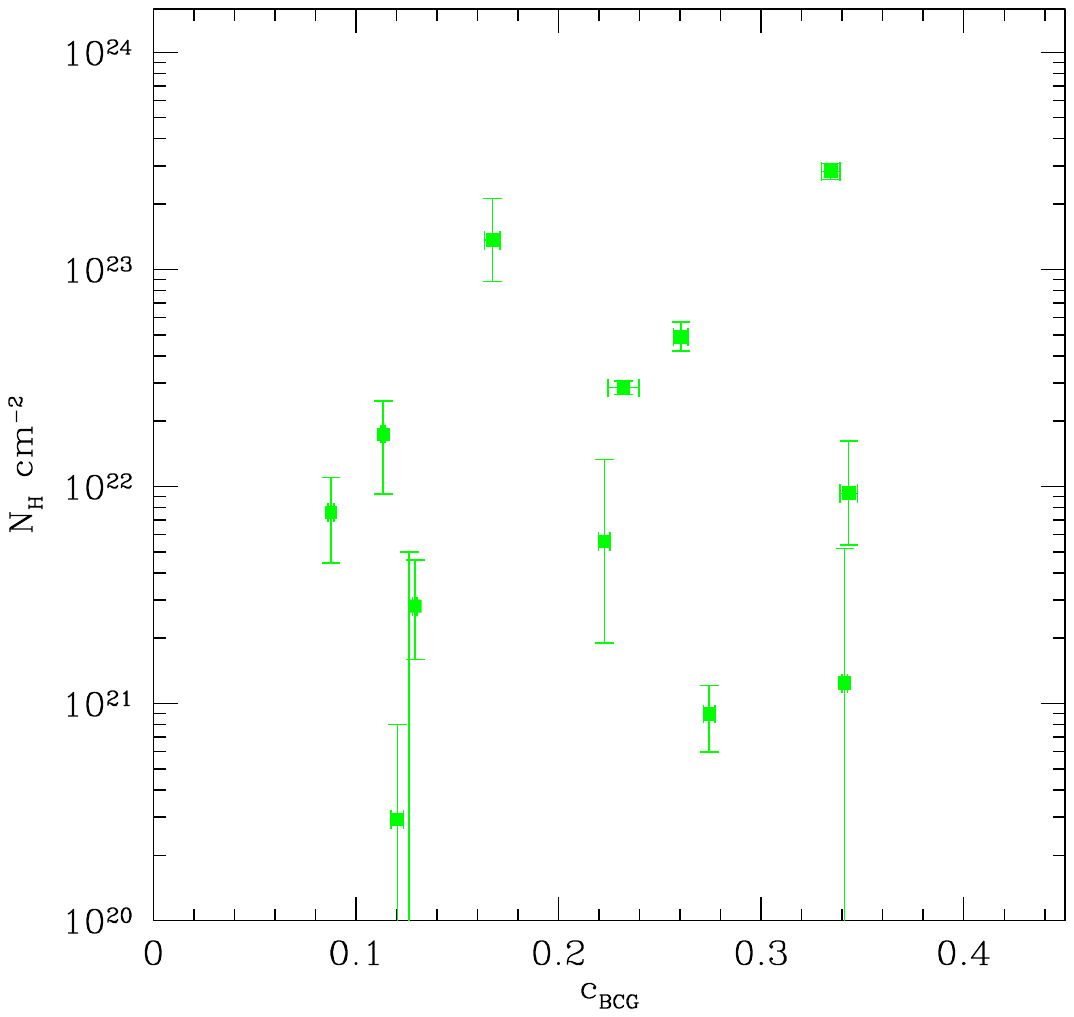} 
\caption{\label{conc_nh}Left:  intrinsic absorption $N_H$ compared to the rest-frame 2-10 keV, unabsorbed luminosity 
as obtained from spectral analysis for the sources in the low-redshift sample.  Right:
intrinsic absorption obtained from spectral fits compared to the concentration parameter at the BCG
position.}
\end{center}
\end{figure}

In the low-redshift bin, we force the spectral analysis on all our sources, including those with low S/N, except for A2125, which is
the source detected with the lowest number of net counts.  The best-fit values of the intrinsic spectral slope, 
intrinsic absorption, and unabsorbed hard-band rest-frame luminosity are shown in  Table \ref{fit_z0.2_0.3}.  
As a simple consistency test, we check that the soft and hard flux values  obtained with our spectral analysis are consistent with 
those obtained with simple aperture photometry within the errors, finding a good agreement.  We find that in general, best-fit 
values for $\Gamma$ range from 1 to 2 with a typical errorbar of 0.25.  In some cases, we find anomalously large or low spectral 
slope (G256.55-65.69, A2146, and CL 2089), showing that for a significant part of our sample, the best-fit values may be driven
by spurious residuals that are due to the direct background subtraction.  We notice that typical values of $\Gamma$ 
for AGN in the Seyfert range of luminosities, are $1.6<\Gamma<2.0$, while our best fit $\Gamma$ are lower on average.  Since we 
are performing spectral analysis in extreme conditions, and small background fluctuations may affect the entire energy range, 
we also perform the spectral analysis by freezing the slope of the power law to $\Gamma=1.8$, which clearly has a significant
effect on the best fit values of the intrinsic absorption. In Figure \ref {fit_vs_cf}, left panel, we compare the values of the 
intrinsic absorption obtained with free power law and with power law frozen to $\Gamma=1.8$.  The largest differences
are obtained for the sources with extremely large or extremely low $\Gamma$, as expected because of the strong
degeneracy between $N_H$ and $\Gamma$.   

In  Figure \ref {fit_vs_cf}, right panel, we investigate whether the unabsorbed luminosities obtained with the spectral analysis are 
consistent with those obtained directly from aperture photometry and our average conversion factors.  We find a good agreement, 
finding that, as expected, the intrinsic absorption of our sources has a modest impact on the luminosity.  Focusing on the 
two sources with $L_X$ discrepant from the values reported in \citet{2013Russell}, we find that the hard luminosity of
RXC J1459 is 1.5 times higher from spectral analysis, which agree with the value found in \citet{2013Russell}.  
However, the hard luminosity from the spectral analysis for A2667 increases, despite the large error bars, and this increases the 
discrepance with respect to \citet{2013Russell}.  Such a difference could be explained only with a 
background three times larger than estimated, which is not acceptable. We note, however, that the hard X-ray emission is displaced
more than 2 arcsec from the peak of the soft emission, and the hard flux may be severely underestimated if the BCG position 
is not firmly secured by the optical image.  

We finally note that all the spectral fits have an acceptable C-statistics, except for two fits.  In the cases of A2146 and CL 2089, we obtain a 
high C-statistics value, and the visual inspection of the residuals shows that this is due to bumps in the low-energy range and at 
the position of the iron emission line complex.  This strongly suggests that a significant contribution from the ICM thermal 
emission has not been 
properly removed by our direct background subtraction.  We also note that these residuals cannot be eliminated by tuning the 
backscale parameter, showing that the problem is not due to a trivial issue of background scaling, but it is related to significant
variation of the thermal properties in the inner 10 kpc.  This aspect can be treated only with a multi-component spectral model, 
an approach that goes beyond the scope of this work.  

In Figure \ref{conc_nh} we present preliminary results related to the distribution of intrinsic absorption. In the left panel, we show the relation between $N_H$ and $L_X$.  We note that the lack of unabsorbed bright ($L_X>10^{43}$ erg s$^{-1}$) 
AGN is significant, while the lack of strongly absorbed, lower luminosity AGN may be due to selection effects against
faint sources.  The statistics is in any case too low to draw any conclusion.  In the right panel of Figure  \ref{conc_nh} we
show the relation between $N_H$ and the ICM concentration parameter, which does not show any obvious trend.

\begin{deluxetable}{cccc} 
\tablewidth{0pt}
\tablecaption{Spectral Analysis of the Sources in the Low-redshift Bin.  
The best-fit parameters are obtained with the model {\tt tbabs ( zwabs $\times$ pow)} with $\Gamma$ free and with 
$\Gamma=1.8$.  $L_X$ corresponds to the unabsorbed, rest-frame 2-10 keV luminosity.  Error bars and upper limits 
corresponds to $1 \sigma$ confidence level. }
\tablehead{  \colhead{Cluster}   &  \colhead{$\Gamma$} & \colhead{ $N_H/10^{22}$ cm$^{-2}$  } & \colhead {$log(L_X$)} }
\startdata
RXC J1504-0248  & 	 $1.59_{-0.22}^{+0.35}$ & $<0.40$ &  $42.56_{-0.05}^{+0.03}$ \\
G256.55-65.69	 & 	 $3.28_{-0.95}^{+1.55}$ & $0.94_{-0.80}^{+1.52}$ &  $42.11_{-0.37}^{+0.30}$ \\
PKS 1353-341	 & 	 $1.17_{-0.14}^{+0.15}$ & $1.63_{-0.31}^{+0.33}$ &  $43.95_{-0.08}^{+-0.03}$ \\
A2390	 & 	 $1.48_{-0.27}^{+0.29}$ & $0.12_{-0.12}^{+0.21}$ &  $42.47_{-0.05}^{+0.03}$ \\
A2667	 & 	 $2.14_{-0.87}^{+2.17}$ & $15.6_{-9.9}^{+15.6}$ &  $43.45_{-0.4}^{+0.47}$ \\
A2146 & 		 $4.48_{-0.16}^{+0.22}$ & $67.7_{-0.06}^{+0.07}$ &  $42.23_{-0.05}^{+0.03}$ \\
RXC J1459.4-1811 & 		 $1.36_{-0.13}^{+0.23}$ & $3.49_{-0.1}^{+0.56}$ &  $43.33_{-0.05}^{+0.03}$ \\
4C+55.16	 & 	 $1.49_{-0.06}^{+0.06}$ & $<0.5$ &  $43.10_{-0.03}^{+0.013}$ \\
CL 2089	 & 	 $-0.78_{-0.09}^{+0.08}$ & $<0.5$ &  $44.10_{-0.3}^{+0.2}$ \\
RXC J1023.8-2715 & 		 $1.13_{-0.23}^{+0.31}$ & $0.04_{-0.03}^{+0.29}$ &  $42.91_{-0.06}^{+0.04}$ \\
CL 0348	 & 	 $1.03_{-0.24}^{+0.28}$ & $0.09_{-0.08}^{+0.28}$ &  $42.91_{-0.07}^{+0.04}$ \\
A611 & 		 $2.03_{-0.21}^{+0.22}$ & $0.11_{-0.09}^{+0.10}$ &  $42.94_{-0.07}^{+0.05}$ \\
3C438	 & 	 $1.00_{-0.16}^{+0.21}$ & $<0.18$ &  $42.63_{-0.07}^{+0.03}$ \\
\hline
RXC J1504-0248	 & 	 $1.80$ & $0.12_{-0.12}^{+0.39}$ &  $42.54_{-0.05}^{+0.02}$ \\
G256.55-65.69	 & 	 $1.80$ & $<0.5$ &  $42.26_{-0.37}^{+0.30}$ \\
PKS 1353-341	 & 	 $1.80$ & $2.86_{-0.19}^{+0.20}$ &  $43.96_{-0.12}^{+0.08}$ \\
A2390	  & 	$1.80$ & $0.28_{-0.12}^{+0.18}$ &  $42.42_{-0.04}^{+0.02}$ \\
A2667	 & 	 $1.80$ & $13.7_{-4.9}^{+7.6}$ &  $43.39_{-0.43}^{+0.22}$ \\
A2146	 & 	 $1.80$ & $1.75_{-0.82}^{+0.72}$ &  $42.25_{-0.10}^{+0.04}$ \\
RXC J1459.4-1811	 & 	 $1.80$ & $4.89_{-0.68}^{+0.85}$ &  $43.42_{-0.19}^{+0.12}$ \\
4C+55.16 & 		 $1.80$ & $0.09_{-0.03}^{+0.03}$ &  $43.02_{-0.03}^{+0.01}$ \\
CL 2089	 & 	 $1.80$ & $28.3_{-2.3}^{+2.5}$ &  $44.50_{-0.50}^{+0.47}$ \\
RXC J1023.8-2715	 & 	 $1.80$ & $0.56_{-0.37}^{+0.78}$ &  $42.84_{-0.06}^{+0.03}$ \\
CL 0348	 & 	 $1.80$ & $0.93_{-0.39}^{+0.69}$ &  $42.85_{-0.07}^{+0.04}$ \\
A611	 & 	 $1.80$ & $0.03_{-0.03}^{+0.05}$ &  $42.99_{-0.04}^{+0.03}$ \\
3C438	 & 	 $1.80$ & $0.76_{-0.31}^{+0.34}$ &  $42.56_{-0.08}^{+0.06}$ \\
\enddata
\label{fit_z0.2_0.3}

\end{deluxetable}

\subsubsection{Spectral Analysis of Sources at $0.55<z<0.75$}


\begin{figure}
\begin{center}
\includegraphics[width=8cm]{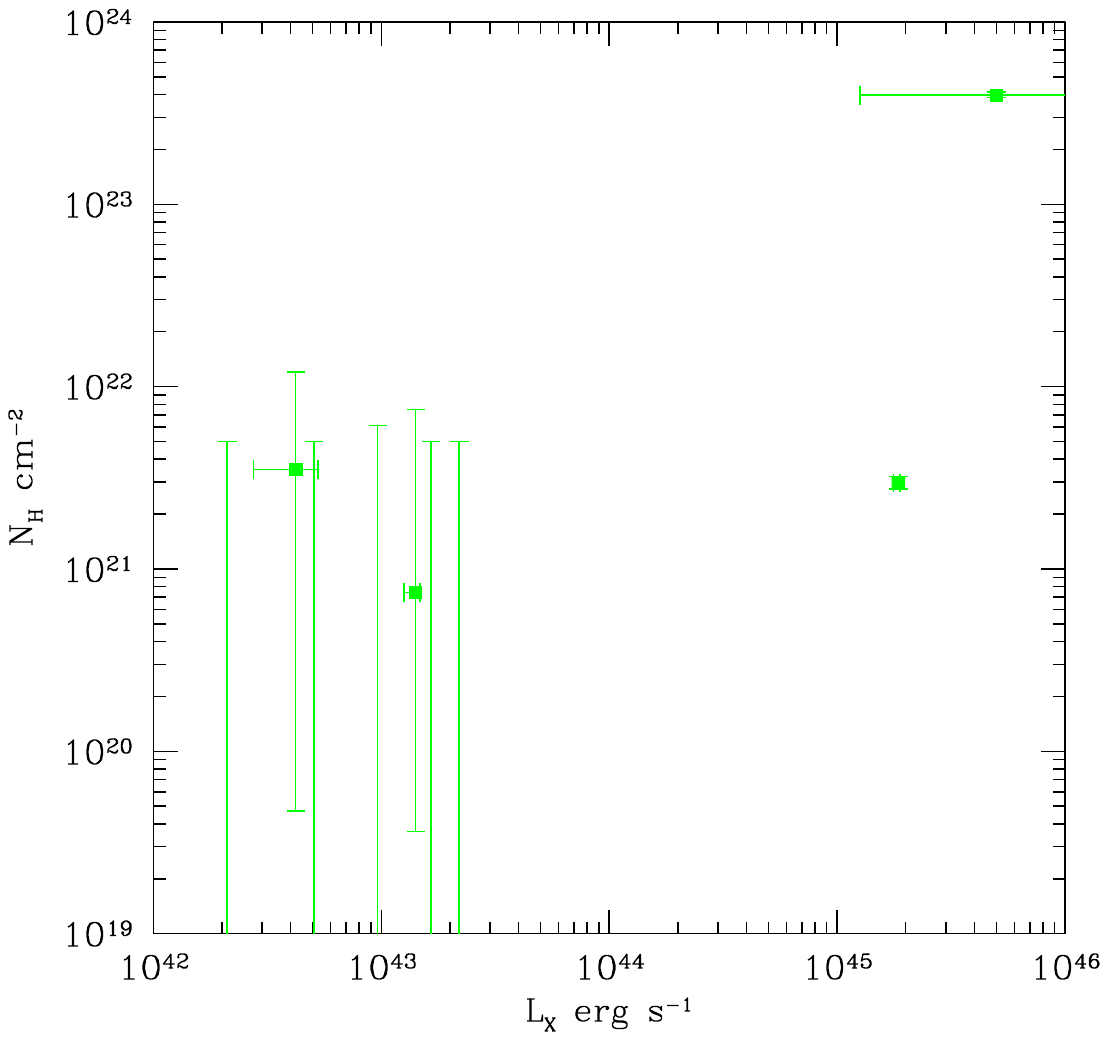}
\includegraphics[width=8cm]{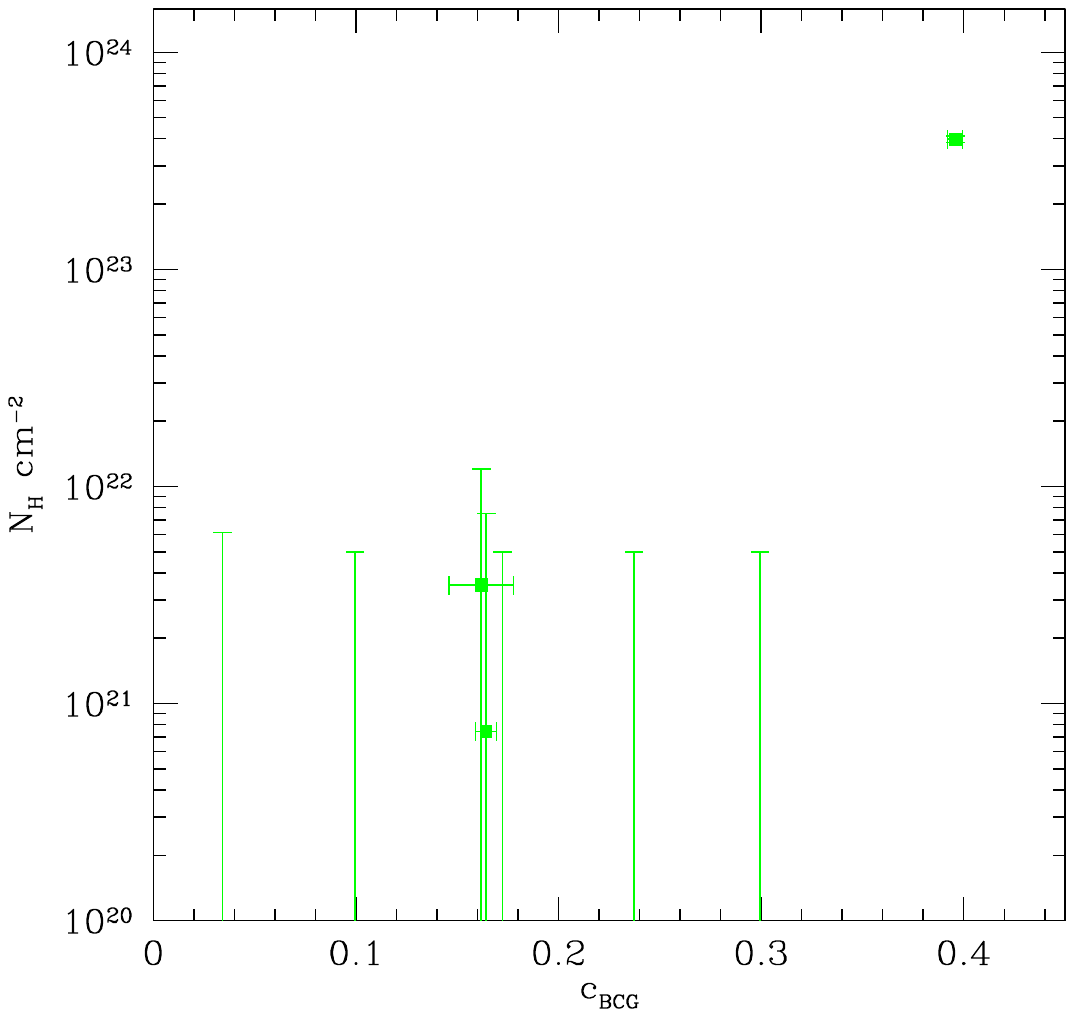} 
\caption{\label{conc_nh_hz}Left: rest-frame 2-10 keV, unabsorbed luminosity obtained from spectral analysis
for the sources in the high-redshift sample, compared to the intrinsic absorption.  Right: 
intrinsic absorption obtained from the spectral fit compared to the concentration parameter at the BCG
position.}
\end{center}
\end{figure}

In the high-redshift bin, we can  perform the fit with the spectral slope $\Gamma$ free only for two sources, finding
again rather flat slopes ($\Gamma \sim 1.2-1.3$).  For all the other sources except for  ACT J0206 and RCS 1107 (which have fewer than 
20 total net counts), we are able to obtain a meaningful spectral fit with spectral slope frozen to $\Gamma=1.8$.  
The results are reported in Table \ref{fit_z0.55_0.75}.  Clearly, the results on $N_H$ are limited with respect to the 
low-redshift bin, since the energy range most sensible to absorption is shifted
out of the observed range.   We are able to confirm that only one source (SPT-CL J2344) has significant absorption, while all
the other sources are consistent with being unabsorbed.  In Figure \ref{conc_nh_hz}, we show the relation 
between $N_H$ and $L_X$ (left panel) and between $N_H$ and $c_{SB}$, which are clearly dominated by upper limits.

\begin{deluxetable}{cccc} 
\tablewidth{0pt}
\tablecaption{Spectral Analysis of the Sources in the High-redshift Bin.  
The best-fit parameters are obtained with the model {\tt tbabs ( zwabs $\times$ pow)} with $\Gamma$ free and with 
$\Gamma=1.8$.  $L_X$ corresponds to the unabsorbed, rest-frame 2-10 keV luminosity.  Error bars and upper limits 
corresponds to $1 \sigma$ confidence level.}
\tablehead{  \colhead{Cluster}   &  \colhead{$\Gamma$} & \colhead{ $N_H/10^{22}$ cm$^{-2}$  } & \colhead {$log(L_X$)} }
\startdata
SPT-CL J2344-4243	 & 	 $1.16_{-0.10}^{+0.10}$ & $27.2_{-2.0}^{+2.2}$ &  $45.49_{-0.39}^{+0.38}$ \\
3C254	 & 	 $1.32_{-0.02}^{+0.02}$ & $<0.04$ &  $45.33_{-0.01}^{+0.01}$ \\
\hline
SPT-CL J2344-4243	 & 	 $1.80$ & $39.8_{-1.3}^{+1.4}$ &  $45.70_{-0.60}^{+0.59}$ \\
RCS 1419+5326	 & 	 $1.80$ & $<0.5$ &  $42.70_{-0.2}^{+0.15}$ \\
SDSS J1004+4112	  & 	$1.80$ & $<0.5$ &  $42.32_{-0.15}^{+0.08}$ \\
MACS 0744.8+3927	 & 	 $1.80$ & $0.07_{-0.06}^{+0.68}$ &  $43.15_{-0.05}^{+0.02}$ \\
SPT-CL J2043-5035	 & 	 $1.80$ & $<0.5$ &  $43.22_{-0.4}^{+0.4}$ \\
3C254	 & 	 $1.80$ & $0.30_{-0.02}^{+0.02}$ &  $45.27_{-0.02}^{+0.01}$ \\
SPT-CL 0001-5748	 & 	 $1.80$ & $<0.5$ &  $43.34_{-0.4}^{+0.4}$ \\
\label{fit_z0.55_0.75}
\enddata

\end{deluxetable}

We conclude that our spectral analysis confirms the results on luminosity and average spectral properties obtained with simple aperture photometry, showing that our approach is effective in studying the X-ray properties of AGN in BCGs.  We also obtain a 
preliminary investigation of the distribution of intrinsic absorption, which is necessarily limited by the statistical error and the 
small number of sources.

\subsection{Comparison of X-ray and Radio Properties}

\begin{figure}
\begin{center}
\includegraphics[width=8cm]{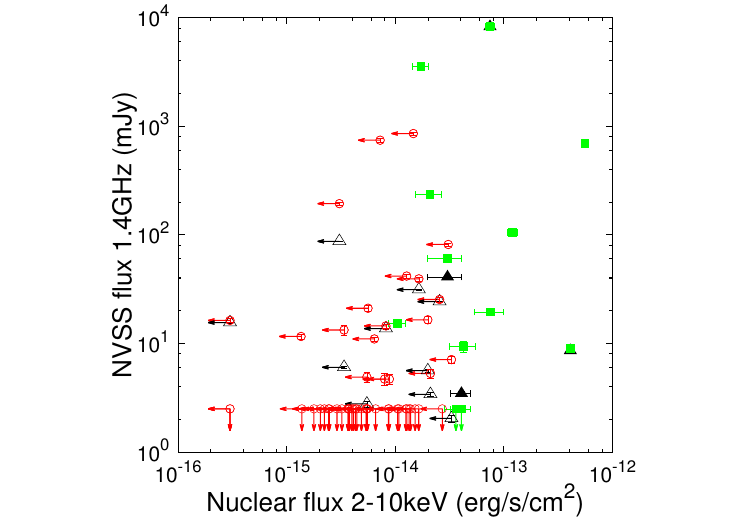}
\includegraphics[width=8cm]{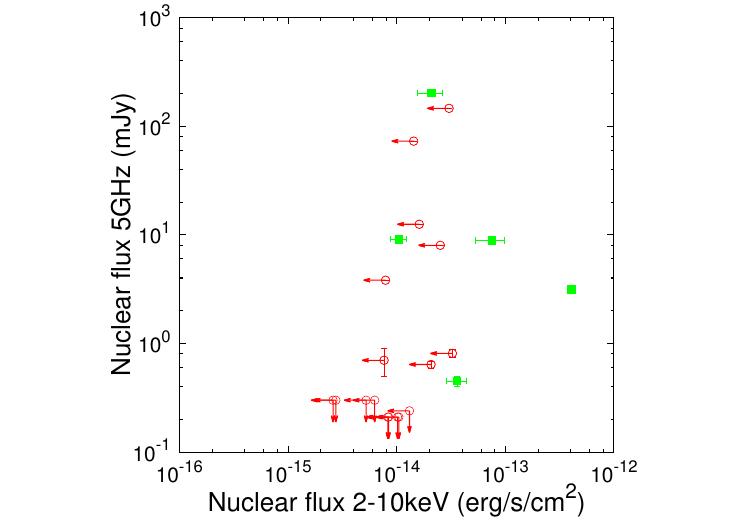}
\caption{\label{XR_NVSS_lowz}Left panel: Radio (1.4\, GHz) and hard X-ray flux scatter plot for BCGs with unresolved 
emission (green solid squares and black triangles) and without unresolved emission (red empty circles) in the redshift range 
$0.2<z<0.3$.  Radio flux is the integrated 1.4~GHz flux from NVSS and FIRST, shown as squares and triangles, respectively. 
Right panel: the same as in the left panel, where the
radio flux is measured at 5.0 GHz by \citet{2015Hogan}.  }
\end{center}
\end{figure}

\begin{figure}
\begin{center}
\includegraphics[width=8cm]{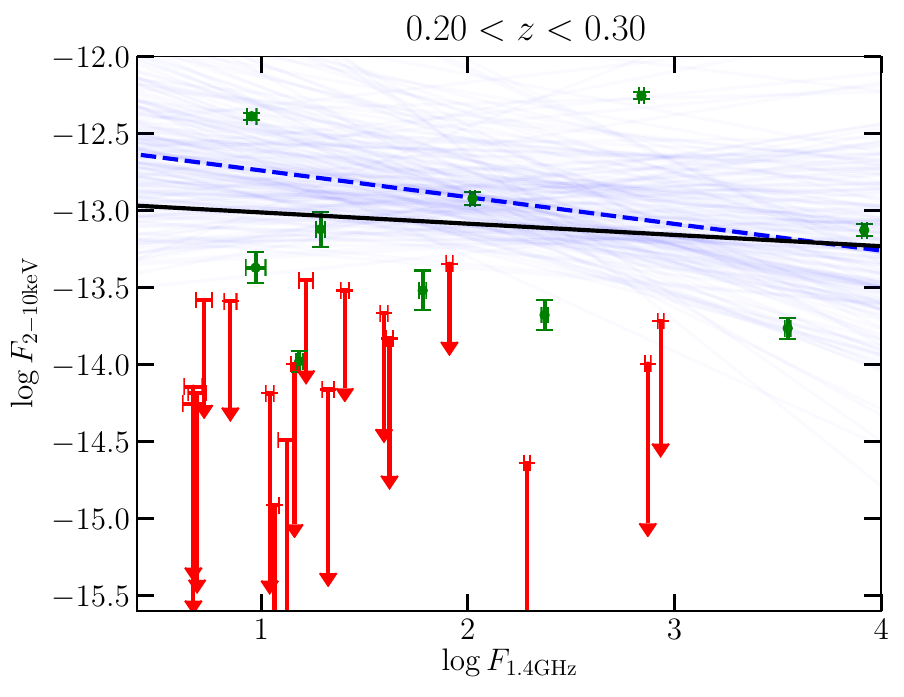}
\caption{\label{frfh}Results from the ASURV regression (solid black line) and the LINMIX\_ERR (dashed blue line)  on the 
correlation between  hard X-ray flux and radio flux density at 1.4\, GHz above 3 mJy.  Light blue lines represent 200 different realizations of the relation from LINMIX\_ERR.}
\end{center}
\end{figure}

\noindent
We also explore the relation between radio and hard X-ray flux in our BCGs.  We identify radio counterparts of
our BCGs in the NRAO VLA Sky Survey \citep[NVSS,\footnote{http://www.cv.nrao.edu/nvss/},][]{1998Condon} 
and Faint Images of the Radio Sky at Twenty-cm \citep[FIRST,\footnote{http://sundog.stsci.edu/}][]{2015Helfand}.
NVSS is complete above $\sim 2.5 \, $mJy at 1.4 GHz for $decl. >-40^{\circ}$, while the FIRST catalog released in 
2014 December  covers about 10,575 square degrees of sky both in the northern and southern hemispheres, 
with the detection threshold  of $\sim  1 $ mJy at 1.4 GHz.   We adopt a simple matching criterion, selecting the NVSS and 
FIRST sources listed in the corresponding catalogs that are closest to the X-ray position of the BCG within a radius of 20 
arcsec and 2 arcsec for NVSS and FIRST, respectively.  A large matching radius is recommended  also for very bright sources in 
NVSS, where 40\% of the FWHM beam size is 20 arcsec, and the FWHM is 45 arcsec\footnote{See R. L. White discussion 
on the NRAO Science Forum https://science.nrao.edu/forums.}.  Since the FIRST resolution is 5.4 arcsec FWHM on average, 
a matching radius of 2 arcsec is chosen for consistency with the 20 arcsec radius used for NVSS sources.
In the low-redshift bin, we identify 29 radio counterparts of our BCGs in NVSS out of 65 sources covered by the survey. 
Of the 65 sources with NVSS data, 13 also have unresolved X-ray emission in the hard band.  For all the other sources with 
NVSS coverage, we assume a conservative upper limit of 2.5 mJy.  We also identify 14 radio counterparts of BCGs out of 29 fields 
covered by FIRST.

In Figure \ref{XR_NVSS_lowz}, left panel, we show the X-ray detected BCGs with green squares, while all the other radio 
counterparts, with only an X-ray flux upper limit, are shown with red circles.  We note that X-ray emission appears at 
any radio power, with a slight preference for low power.  In any case, there are no hints of a correlation between 
hard X-ray and radio emission from BCGs in the $0.2<z<0.3$ redshift range.  In the right panel of Figure 
\ref{XR_NVSS_lowz} we also show the scatter plot of the X-ray  and 5 GHz radio flux for the 22 sources in common
with the sample studied by \citet{2015Hogan}.   A visual inspection of Figure \ref{XR_NVSS_lowz} shows that there 
are no clear signs of a correlation between the hard flux $F_H$ and the radio flux density $F_R$ both at 1.4 GHz and 5 GHz (left 
and right panel, respectively).  A censored-data analysis is very challenging because of the many double upper limits. 
If we search for a correlation for radio flux densities above $\sim 3$ mJy at 1.4 GHz (just above the completeness level of the 
NVSS), we are able to obtain a best fit with LINMIX\_ERR and ASURV.  In both cases we find a slope consistent with
zero and therefore no signs of correlation (see Figure \ref{frfh}). 
Russell et al. 2013 (see their section 3.6) did not find a correlation between the nuclear radio 5 GHz and X-ray fluxes either. 
{\bf The absence of a radio correlation suggests that massive sub-relativistic outflows may be the primary driver of 
kinetic feedback, instead of relativistic jets.}

Finally, we note a few cases where a radio source is present in the NVSS field of view close to the 
BCG, but is not listed in the NVSS catalog, and therefore is not included in our preliminary cross correlation 
between our BCG and radio counterpart.  We stress that a high-resolution follow-up of our BCG with JVLA is needed to 
firmly identify counterparts of our BCG and to exclude interlopers or non-BCG cluster members, as shown in a few cases in 
our program of JVLA observation of BCG in the CLASH sample \citep{2018Yu}.


\section{Discussion: Implications for AGN Feeding and Feedback}\label{s:disc}

\noindent
Keeping in mind the limited statistics, we discuss here some implications for the accretion and feedback mode tied to the SMBHs at 
the center of BCGs and the associated phenomenology in the X-ray and radio bands.  As introduced in \S\ref{s:intro}, the 
maintenance mode of AGN feedback occurs via mechanical injection of energy (\citealt{McNamara:2012} for a review). Ultrafast 
AGN outflows and/or relativistic jets are launched within the inner 100 gravitational radii from the SMBH, as shown by high-quality 
X-ray data \citep[e.g.][]{2015Nardini} and confirmed by general-relativistic magnetohydrodynamics (MHD) simulations (see \citealt{Sadowski:2017} and 
references within).  Such simulations imply that below a few percent of the Eddington rate, the radiative power is expected to be 
fewer than the kinetic input (see also \citealt{2013Russell}).  Consistently with this picture, we find that less than 20\% of BCGs 
are X-ray bright with a cutoff near $10^{43}$ erg s$^{-1}$ in the low-z bin. Therefore, only a handful of sources appear to 
approach the radiatively efficient regime. We thus expect 
mechanical feedback to dominate over radiative feedback (radiation pressure or Compton heating) also in our sample. 
Note {\bf that the total power} AGN outburst in clusters can reach $\sim10^{45}$\,erg\,s$^{-1}$, as observed
\citep{2015Hlavacek-Larrondo} and predicted by simulations (e.g., \citealt{2012Gaspari}), thereby our nuclear X-ray 
luminosities may be 100\,-\,1000 times lower than the maximal injected AGN power.  {\bf In this framework, the investigation
of the nuclear luminosity of BCG at higher redshift may be key to constrain the switching of the feedback mode from 
mechanical to radiatively efficient, as has been suggested by \citet{2013aHlavacek-Larrondo} for clusters with 
clear cavities in the ICM.}  Radiative feedback is indeed expected to increase at higher 
redshift due to the halos are progressively smaller and SMBH masses are progressively smaller, and hence  Eddington ratios are larger.

Our {\bf approach} may provide further constraints to the feedback mechanism.  For example, the absence of evolution in the 
bulk of the population at moderate luminosities ($L_X<10^{44}$ erg s$^{-1}$), {\bf if found 
in a larger sample of virialized halos with no obvious selection bias and on a wider redshift range extending beyond $z=1$}, 
would imply that mechanical AGN feedback is tightly self-regulated since at least $\sim$ 7 Gyr, {\bf regardless of the dynamical 
state and age of the halo}.  This would be in agreement with the presence of cool cores up to redshift 1.9
(e.g., \citealt{2017McDonald}).  

Regarding feeding, the likely source of accretion onto the the SMBH comes from the significant amount of cooling gas out 
of the hot plasma filling BCGs, groups, and clusters, {\bf as suggested by the fact that
within a few tens of kilparsec, the cooling time typically becomes much lower than $\sim 100$ Myr.} 
Turbulent motions (driven by AGN outbursts and mergers; e.g., \citealt{2016Hitomi}) trigger nonlinear thermal instability, 
promoting the condensation of warm ($10^4$\,K) filaments and cold ($<50$\,K) clouds in a top-down multiphase condensation 
cascade, a scenario that has been probed with multiwavelength thermodynamic \citep[e.g.][]{Gaspari:2017_uni} and kinematic 
\citep[e.g.][]{2018Gaspari} tracers.  During CCA, the clouds collide inelastically within $r< 500$\,pc, promoting rapid radial 
funneling down to a few tens gravitational radii, hence rapidly boosting the accretion rate,  
without the requirement of a thin disk.  In addition, shells of gas lifted by powerful AGN outflows 
may fragment through Rayleigh-Taylor instabilities and produce clouds of cooling gas that may eventually fall back 
toward the black hole and contribute to its feeding \citep[e.g.,][]{2017Gilli}.  {\bf This process affects both the 
shape of the average nuclear luminosity of the BCG and its variance.  In particular,} a flicker noise 
variability is expected to have a power spectrum logarithmic slope of -1, characteristic of fractal and chaotic phenomena. 
{\bf At the same time, while on average, the 
rates from the clumpy rain in BCGs are expected to remain significantly sub-Eddington, variations of $\sim 2$ dex are 
expected. Our investigation, applied to a larger sample, will provide significant constraints
on these two observables, hence on the accretion mechanism.}

Another scenario for the presence of X-ray emission may be related to a relatively stable, classic thin accretion disk.
While in CCA an accreting structure may develop within tens gravitational
radii (similar to a thick torus), the clumpy nature of the rain onto the SMBH makes it 
difficult for the disk to survive intact. Once again, the key difference is the strong and rapid variability (flicker noise) 
induced by the continuously raining clouds, {\bf an aspect that can be investigated by exploiting the full {\sl Chandra} archive.} 

{\bf Finally, an important aspect of our approach is the full spectral analysis of the X-ray emitting BCGs.  The 
intrinsic X-ray absorption, together with obscuration properties in other bands, such as UV and optical, and even in the molecular 
regime \citep[see the case of A2597 in][]{Tremblay:2016} may be used to constrain the clumpiness, which is predicted to appear 
since the initial stages of the condensation cascade in the CCA scenario.  Indeed, the multiwavelength studies of 
residual cooling in and around BCGs are a crucial testbed of the primary feeding mechanism 
(e.g., \citealt{2011McDonalda,2014Werner,Tremblay:2015,Voit:2015_nat,Hamer:2016}).}

\section{Conclusions}
\noindent
We measured the cumulative fraction of X-ray luminous BCGs as a function of the 2-10 keV luminosity 
in the redshift ranges $0.2<z<0.3$ and  $0.55<z<0.75$. We compiled together our BCG sample without any preliminary 
selection on the host clusters, simply collecting all the available observations of clusters and groups public as of 2016 September 
with more than 20 ks of total exposure in the {\sl Chandra} archive. Our aim is to constrain the history of accretion onto 
the SMBH of BCGs galaxies across cosmic epochs, and ultimately, its effect on the feedback duty cycle. This last piece of 
information is clearly a key ingredient toward the comprehension of the baryonic cycle at the center of groups and 
clusters of galaxies.  In this preliminary work, we investigated the presence of X-ray nuclear emission in the BCGs, whose
measurement is made difficult by 
the overwhelming emission of the surrounding ICM, particularly in cool-core clusters.  However, thanks to the exquisite 
angular resolution of {\sl Chandra}, and the use of the hard band, where the ICM emission is lower and the AGN emission 
less affected by intrinsic absorption, we were able to probe the presence of X-ray nuclear activity down to luminosities as 
low as $10^{42}$ erg s$^{-1}$. Our results can be summarized as follows: 

\begin{itemize}
\item about 18\% (14 out of 81 and 9 out of 51 in the low- and high-z sample, respectively) of the BCGs show unresolved 
X-ray emission in the 0.5-2 keV or 2-7 keV bands, a fraction that is significantly lower than that found in clusters with large 
X-ray cavities by \citet{2013Russell};
\item some of the X-ray emitting BCGs (at least half in the low-z bin) appear to have significant intrinsic absorption on the 
basis of their hardness ratio in the $0.2<z<0.3$ redshift range;
\item in the low-redshift sample, hard X-ray luminosities range from $1.6\times 10^{42}$ to 
$\sim 10^{44}$ erg s$^{-1}$ and the cumulative fraction has a slope between $\sim$\,-0.6 
and $\sim$\, -1, with a weak hint of a steeper slope at $L_X\geq 10^{43}$  erg s$^{-1}$;
\item after accounting for the flux limits of our detections, we find no evidence for evolution {\bf in our sample} at luminosities
$L_X<10^{44}$ erg s$^{-1}$ between $\langle z\rangle \sim 0.25$ and $\langle z\rangle \sim 0.65$;
\item the only two sources with quasar-like luminosity  ($L_X>10^{45}$ erg s$^{-1}$) are both in the $0.55<z<0.75$ range;
\item X-ray spectral analysis shows that hard-band luminosities based on photometry are robust, and confirms the presence of
significant intrinsic absorption $N_H>10^{22}$ cm$^{-2}$ for about half of the sample in the low-redshift bin;
\item the correlation with the BCG concentration parameter $c_{BCG}$ shows that X-ray luminous BCGs 
($L_X>10^{42}-10^{43}$ erg s$^{-1}$ in the low- and high-redshift bin, respectively) tend to 
be in bright cores, although most of the strongest cores do not host nuclear X-ray emission;
\item we do not find any significant correlation between X-ray luminosity and radio power;
\item the low nuclear luminosities suggest that the main mode of feedback, even in X-ray bright BCGs, is mechanical and not 
radiatively driven; the absence of a radio correlation suggests that massive sub-relativistic outflows may be the primary driver of 
kinetic feedback, instead of relativistic jets;
\item {\bf the percentage of outliers with high luminosities and the measurement of intrinsic absorption in soft X-rays, as 
envisaged in our approach, can efficiently complement other multiwavelength BCG studies to constrain the primary channel of 
the SMBH feeding such as CCA.}
\end{itemize}

The results summarized here must be considered as preliminary, since the sample selection, based simply on the 
public observations of groups clusters in the {\sl Chandra} archive, does not guarantee the control of possible bias.
{\bf On the other hand, an unbiased sample of virialized halos can be obtained only by combining observations of
 X-ray, SZ, optical, and radio-selected groups and clusters already available in the {\sl Chandra} archive.}
Therefore, we plan to extend our analysis to the largest possible dataset, and eventually extract subsamples of targets 
with different selection function {\bf to quantify the effects of selection bias on our observables}. 
As a next step, we will relax the constraints on the redshift range and on the 
minimum exposure time, and collect multiwavelength data to complement X-ray with 
measurements of the SMBH mass, mass of the host halo, presence of cavities, dynamical state of the halo, and 
spectral characterization of the cool-core strength.  Our final goal is to investigate the origin of the feeding gas and 
the accretion regime in BCGs at different cosmic epochs as a function of the environment.

\acknowledgments
We thank Marco Chiaberge for useful discussions. P.T. is supported by the Recruitment Program of 
High-end Foreign Experts, and he gratefully acknowledges the hospitality of Beijing Normal University.
H.Y. is supported by the National Natural Science Foundation of China 
under Grants Nos. 11403002 and the Fundamental Research Funds for the 
Central Universities.
E.L. is supported by a European Union COFUND/Durham Junior Research Fellowship (under EU grant agreement number 
609412).  M.G. is supported by NASA through Einstein Postdoctoral Fellowship Award Number PF5-160137 issued by the 
{\sl Chandra X-ray Observatory} Center, which is operated by the SAO for and on behalf of NASA under contract NAS8-03060. 
Support for this work was also provided by Chandra grant GO7-18121X. E.N. acknowledges funding from the European 
Union's Horizon 2020 research and innovation programme under the Marie Sk\l{}odowska-Curie grant agreement No. 
664931.

\providecommand{\SortNoop}[1]{}

\end{document}